\documentclass[aps,prb,twocolumn,showpacs,floatfix,superscriptaddress,byrevtex]{revtex4-1}
\usepackage{graphicx}
\usepackage{graphics}
\usepackage{epsfig}
\usepackage{amsmath}
\usepackage{amsfonts}
\usepackage{amssymb}
\usepackage{dcolumn}
\usepackage{bm}
\usepackage{longtable}
\usepackage{multirow}

\newcommand{\ee} {\text{e}}
\newcommand{\ii} {\text{i}}
\newcommand{\vctr}{{\bf{r}}}
\newcommand{\vctR}{{\bf{R}}}

\newcommand{\NT}{\text{NT}}

\begin{document}

\title{Inter-valley scattering induced by Coulomb interaction and disorder
in carbon-nanotube quantum dots}

\author{Andrea Secchi}
\email{andrea.secchi@gmail.com}
\altaffiliation[Present address: ]{Institute for Molecules and Materials,
Radboud University Nijmegen, The Netherlands}
\affiliation{CNR-NANO S3, Via Campi 213a, 41125 Modena, Italy}
\affiliation{Universit\`a degli Studi di Modena and Reggio Emilia, Italy}

\author{Massimo Rontani}
\email{massimo.rontani@nano.cnr.it}
%\homepage{www.nanoscience.unimore.it/max.html}
\affiliation{CNR-NANO S3, Via Campi 213a, 41125 Modena, Italy}

\date{\today }

\begin{abstract}
We develop a theory of inter-valley Coulomb scattering in 
semiconducting carbon-nanotube quantum dots, taking into account the effects 
of curvature and chirality. Starting from the effective-mass
description of single-particle states, we study the two-electron system 
by fully including Coulomb interaction, spin-orbit coupling, and short-range 
disorder. We find that the energy level splittings associated with 
inter-valley scattering are nearly independent of the chiral angle and, 
while smaller than those due to spin-orbit interaction, large 
enough to be measurable.
\end{abstract}


\maketitle

\section{Introduction}

Carbon nanotubes \cite{DresselhausBook, AndoReview} (CNTs) have emerged in the 
last two decades as ideal realizations of one-dimensional (1D) quantum systems. Indeed, for electronic excitations 
close enough to the charge neutrality 
point, the longitudinal degrees of freedom are effectively decoupled from 
the transverse ones.\cite{IlaniAnnRev, DeshpandeRev, Kuemmeth10} 
Advances in employing as-grown suspended CNTs 
as Coulomb-blockade devices\cite{Cao05} 
allowed for dramatically reducing the disorder in the samples 
and the dielectric screening due to the environment. 
This breakthrough led to the recent observation of 
fascinating many-body states, such as the Wigner 
molecule\cite{Deshpande08,Pecker13}---the finite-size analog of
the Wigner crystal---and the Mott-Hubbard insulator,\cite{Deshpande09} 
as well as to the measurement of significant
spin-orbit 
coupling.\cite{Kuemmeth08,Churchill09,Jhang10,Jespersen11,Steele13} 
The latter is enhanced with respect to graphene because of the 
curved topology  of the CNT 
surface.\cite{Ando00,HuertasHernando06,Zhou09,Chico09,Jeong09,Izumida09} 
An important feature of CNTs is the absence of hyperfine
interaction since C nuclei have zero spin.\cite{Bulaev08}
This has fueled the pursuit of spin qubits in CNT quantum dots 
(QDs).\cite{Buitelaar08,Churchill09,Churchill09Nat,Steele09,Palyi10,vonStecher10,Chorley11,Coish11,Reynoso11,Palyi11,Reynoso12,Pei12}
Interestingly, spin-orbit interaction in CNTs
may be useful for spintronics and
quantum-information purposes.
In fact, one could manipulate spins by means of either
electric fields acting on the orbital degrees of
freedom\cite{Palyi10,Jespersen11,Klinovaja11} or by exploiting
bends,\cite{Flensberg10} or even encode information in the
valley index.\cite{Palyi11}

A remarkable property that  
distinguishes CNTs from other 1D devices is the occurrence of two 
spinorial degrees of freedom, one being the real electron 
spin $\sigma = \pm 1$, 
the other one being the \emph{isospin} $\tau = \pm 1$ associated with the
valley population in reciprocal space. The latter is well defined 
close to the two 
non-equivalent points K and K$'$ at the border of the Brillouin zone,
where the apices of graphene's Dirac cones touch. 
The isospin is commonly assumed to be a good quantum number,
which is true for electrons scattered by potentials that are
slowly varying in space with respect to the graphene lattice constant $a$ 
(with $a = 2.46$ {\AA}).\cite{Wallace47,Luttinger55,Ando92}
However, if the momentum transferred during scattering is $\sim 1/a$, it
may make electrons to swap valleys, as  
the distance between the two valleys in momentum space is 
$\sim\left| {\bf{K}} - {\bf{K}}' \right| = 4 \pi / (3 a)$. 
The object of this Article is the theory of inter-valley scattering. 
We are mainly interested in  
the role of 
inter-valley scattering in Coulomb blockade experiments,
hence we focus on gate-defined QDs embedded 
in semiconducting CNTs in the few-electron regime. 

So far, the vaste majority of analytical or semi-analytical theories 
based on the envelope function in the effective mass
approximation\cite{Luttinger55}
has regarded inter-valley scattering as being either negligible or  
small with respect to other sources of 
scattering.\cite{Ando06,Secchi09,Wunsch09,Secchi10,Weiss10,Secchi12,Roy12}
This should not come as a surprise, since inter-valley scattering is
inherently not included in the envelope function theory, 
with the envelope being built as a superposition of Bloch
states whose wave vectors lie close to the bottom of one valley.
A few theories have considered
the scattering induced by short-range disorder,
such as atomic scale defects.\cite{Palyi10,Rudner10}
A channel of inter-valley scattering of special interest here is 
the one induced by the short-range part of Coulomb interaction, also known 
as backward (BW) scattering.\cite{Egger97,Egger98,Ando06,Mayrhofer08} 
Indeed, BW Coulomb interaction exchanges the isospins of two 
electrons, since these degrees of freedom are ultimately related to 
the orbital component of the wave function: electrons with different 
isospins have different crystal momenta, and hence 
different microscopic Bloch states. With respect to the long-range part
of Coulomb interaction, conserving valley quantum numbers  
[known as intra-valley, or forward (FW) scattering], the
BW scattering term is much weaker.\cite{Secchi09,Wunsch09,Secchi10,Weiss10}
Note that both FW and BW terms conserve the total crystal momentum
in the scattering event.

On the experimental side, growing evidence shows that inter-valley 
scattering is significant and measurable. Low-temperature transport data 
rely on Coulomb blockade spectroscopy,
based on the precise control of the electron
number in CNT QDs down to the single electron.\cite{Reimann02,Hanson07} 
One connects source and drain electrodes  
to a CNT and operates on a 
capacitatively coupled electrostatic gate, allowing to rigidly
shift the QD energy spectrum with respect to Fermi energies of the
leads. If the QD chemical potential falls outside the 
transport energy window controlled by the source-drain bias, 
no current flows and the electron number $N$ in the dot is fixed. 
Otherwise, electrons 
may tunnel from the source to the drain trough the QD while its 
population fluctuates between $N$ and $N+1$. By recording the 
differential conductance as a function of the source-drain bias 
and gate voltage one measures the evolution
of ground- and excited-state chemical potentials vs the external
magnetic field, linking the slopes of the curves to (iso)spin quantum 
numbers.\cite{Minot04,Jarillo-Herrero04,Cao05,Deshpande08,Kuemmeth08,Churchill09,Steele09,Churchill09Nat,Jespersen11,Pei12,Pecker13,Steele13} 
This spectroscopy allowed to clearly resolve the
anticrossings between energy levels of opposite valleys, 
that were attributed to short-range disorder.\cite{Kuemmeth08,Jespersen11} 
Besides, the recent observation of a two-electron Wigner molecule in a 
CNT QD pointed out the significant role of BW scattering in the fine structure
of the low-lying excited states, inducing energy splittings
comparable to those associated with spin-orbit interaction.\cite{Pecker13}

Moreover, the quantitative determination of BW interaction is 
important for studies of Pauli spin and valley 
blockade in coupled QDs,\cite{Buitelaar08,Churchill09,Churchill09Nat,Palyi10,vonStecher10,Chorley11,Palyi11,Coish11,Reynoso11,Reynoso12,Pei12} 
aiming to realize spin-to-charge 
conversion useful for applications.
If $(n_L,n_R)$ are the electronic populations of the left and right QD, 
respectively, the resonant tunneling sequence is the cycle
$ (0,1) \rightarrow (1,1) \rightarrow (0, 2) \rightarrow (0,1)$,
where the left (right) dot is the one close to the source (drain) electrode. 
A given intermediate state $(0, 2)$ is Pauli-excluded from 
transport if its total (iso)spin is incompatible with the projection $\sigma$ 
($\tau$) carried by the tunneling 
electron.\cite{Weimann95,Ono02,Churchill09,Pecker13}
When the two electrons come close to each other in the right dot BW 
scattering becomes relevant, mixing the eigenstates of (iso)spin and
hence relaxing the Pauli blockade.

Whereas optical properties of CNTs are beyond the scope ot this work,
we mention that BW interaction crucially dictates the fine
structure of excitons, controlling the sequence of 
bright and dark excitons as well as their energy 
splittings.\cite{Chang04,Perebeinos04,Zhao04,Maultzsch05,Wang05,Zaric06,Ando06,Mortimer07,Shaver07,Jiang07,Srivastava08,Matsunaga08,Torrens08}
Computational approaches based on the full numerical
solution of the Bethe-Salpeter equation were
applied to the smaller CNTs\cite{Chang04,Spataru04} together with 
simpler but more transparent theories for larger tubes, such as semi-empirical 
models\cite{Zhao04,Aryanpour12} as well as 
treatments within the effective-mass approximation\cite{Ando06}
or the tight-binding method.\cite{Perebeinos04,Jiang07,Goupalov11}
Few experimental data are available 
since dark excitons are optically inactive and hence difficult 
to observe.\cite{Maultzsch05,Wang05,Zaric06,Mortimer07,Shaver07,Srivastava08,Matsunaga08,Torrens08}

The main goal of this Article is the analysis of the impact 
of BW scattering on carbon-nanotube quantum dots. 
We model the gate-defined QD as a 1D harmonic trap, using  
sublattice envelope functions in the 
effective mass approximation. 
The exact diagonalization\cite{Rontani06} 
of the long-range part of Coulomb interaction for two electrons 
fully takes into account 
spin-orbit (SO) coupling, BW scattering, and disorder---in the 
form of a generic distribution of defects.
In our detailed investigation we consider
the dependence of BW interaction on the microscopic CNT structure
(i.e., on the chiral angle $\alpha$ in addition to the radius $R$), 
going beyond the previous 
treatment of BW interaction as a contact 
force.\cite{Ando06,Secchi09,Wunsch09,Secchi10,Secchi12}
We include BW scattering and defects at the level of 
first-order perturbation theory. Specifically,
we present analytical expressions for the energies and the spin-isospin part 
of the two-electron wave function. Such expressions depend only on 
two parameters, related respectively to the orbital component
of the wave function, 
which can be evaluated through exact 
diagonalization,\cite{Rontani06,Secchi09,Secchi10,Secchi12,Pecker13} 
and the distribution of disorder. We estimate that energy splittings 
due to BW scattering may be about one order of magnitude smaller than 
those induced by SO interaction but large enough to be measurable 
in experiments. 

The short-range BW interaction  
is sensitive to
the relative position of two electrons in the QD, which is
controlled in turn
by the competing effects of the long-range part of Coulomb interaction 
and confinement potential. 
Whereas Coulomb repulsion tends to push electrons aside, 
the QD confinement potential squeezes them towards the QD center. 
When Coulomb energy overcomes the sum of kinetic and confinement energy, 
electrons localize in space \emph{\`a la} Wigner,
arranging themselves in a geometrical configuration [a Wigner molecule (WM)] 
to minimize the electrostatic energy. Signatures of Wigner 
crystallization were predicted theoretically\cite{Secchi09,Wunsch09,Secchi10,Roy10,Secchi12,Ziani12,Mantelli12} and 
observed experimentally.\cite{Deshpande08,Pecker13} 
Note that, as a consequence of localization, exchange interactions 
are suppressed, hence states with the same charge density and different 
(iso)spin projections become degenerate. 

The fact that the energy cost 
needed to flip the (iso)spin is tiny makes the WM regime
detrimental for device operations based on Pauli blockade. 
Therefore, in this Article we also consider the opposite, weakly interacting
regime where confinement energy overcomes the Coulomb energy. This may be
achieved if: \cite{Secchi10} (i) the QD is sufficiently small 
(ii) $R$ is large (iii) the effective dielectric screening
due to the presence of leads and gates is significant. 
Our results show that the impact of the BW contact interaction
increases going from the WM to the weakly interacting regime, consistently
with the shrinking of the correlation hole.

This Article is organized as follows. 
In Sec.~\ref{s:coordinates} we work out the coordinates of the atoms
of a generic CNT in a frame oriented along the tube axis, which 
we later use to evaluate the BW scattering potential. 
After introducing the many-body Hamiltonian (Sec.~\ref{s:MB}), 
Sec.~\ref{s:BW} provides
an exhaustive discussion of BW scattering. 
In Section \ref{s:two} we recall 
from Refs.~\onlinecite{Secchi09,Secchi10} the results
on the two-electron system in the absence
of SO coupling, BW scattering, and disorder. 
We include BW scattering and SO interaction in 
Sec.~\ref{s:two_plus_BW}, and then compare our predictions
with the available experimental results (Sec.~\ref{s:exp}). 
The final step is to include short-range disorder in the theory
(Sec.~\ref{s:two_plus_disorder}). 
After the Conclusion (Sec.~\ref{s:conclusion}), 
in the Appendixes we present the details of the derivation of
atomic coordinates (App.~\ref{a:coordinates}),
the properties of the single-particle basis set (App.~\ref{s:SP}),  
the BW term of the Hamiltonian (App.~\ref{a:BW}),
and the form of the Hamiltonian in the disordered case
(App.~\ref{a:disorder}).

\section{Atomic coordinates of carbon nanotubes}\label{s:coordinates} 

In this section we determine the cylindrical coordinates 
of the carbon atoms of the CNT orienting the vertical coordinate
along the tube axis $y$. This task,
which is not trivial for a generic CNT, is needed to 
subsequently include the
effects of curvature and chirality into
the BW term of the Hamiltonian (cf. Sec.~\ref{s:BW}). 

\begin{figure}
%\vspace{4mm}
\centerline{\epsfig{file=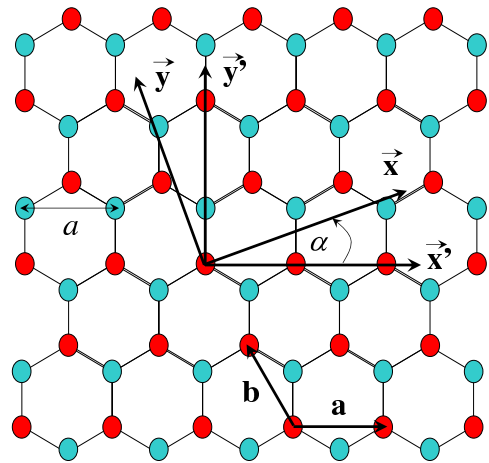,width=2.5in,angle=0}}
\caption{(Color online) Schematic representation of the graphene lattice and 
two useful reference frames. 
Here $\alpha$ is the CNT chiral angle. 
The $y$ axis is parallel to the nanotube axis, 
whereas the $x$ axis is parallel to the chiral vector.  
\label{chiral}}
\end{figure}

In graphene, the atomic coordinates for sublattices A and B,
respectively ${\bf{R}}_{\text{A}}$ and ${\bf{R}}_{\text{B}}$, may be 
written as
\begin{align}
& {\bf{R}}_{\text{A}} (n_1, n_2) =  a \! \left[ \left( n_1 - \frac{1}{2} n_2 \right) \overrightarrow{\bf{x}}' + \left( \! \frac{\sqrt{3}}{2} n_2 + \frac{1}{\sqrt{3}} \! \right) \overrightarrow{\bf{y}}'\right] , \nonumber \\
& {\bf{R}}_{\text{B}} (n_1, n_2) = a \! \left[ \left( n_1 - \frac{1}{2} n_2 \right) \overrightarrow{\bf{x}}' + \frac{\sqrt{3}}{2} n_2 \overrightarrow{\bf{y}}'\right] ,
\label{coordinates graphene}
\end{align} 
where $n_1$ and $n_2$ are integers, $a = 2.46$ {\AA} is the lattice 
parameter of graphene, and the unit vectors 
$\overrightarrow{\bf{x}}'$ and $\overrightarrow{\bf{y}}'$ are shown 
in Fig.~\ref{chiral}. The two atoms A and B specified by the same 
couple of integers  $(n_1, n_2)$ belong to the same graphene unit cell. 
Folowing the well known procedure\cite{DresselhausBook,AndoReview}
of wrapping the graphene sheet to form
the CNT,
we define the chiral vector of the CNT,
connecting now equivalent sites of the tube, 
as ${\bf{L}} \equiv n_a {\bf{a}} + n_b {\bf{b}}$, where 
\begin{align*}
{\bf{a}} = a \overrightarrow{\bf{x}}', \quad \quad 
{\bf{b}} = a \left( -\frac{1}{2} \overrightarrow{\bf{x}}' + \frac{\sqrt{3}}{2} \overrightarrow{\bf{y}}' \right)
\end{align*}
form a basis for the graphene lattice; the length of the chiral vector is $L = a \sqrt{n^2_a + n^2_b - n_a n_b}$. The chiral angle $\alpha$ is the angle between ${\bf{L}}$ and the unit vector $\overrightarrow{\bf{x}}'$; because
of the hexagonal symmetry, it can be always chosen to lie in the interval $[0, \pi/6]$. It is determined by
\begin{align}
& \cos(\alpha) = \frac{{\bf{L}} \cdot \overrightarrow{\bf{x}}'}{L} = \frac{n_a - n_b/2}{\sqrt{n^2_a + n^2_b - n_a n_b}}, \nonumber \\ 
& \sin(\alpha) = \frac{\sqrt{3} n_b}{2 \sqrt{n^2_a + n^2_b - n_a n_b}}.
\label{alpha}
\end{align}
We now rotate the reference frame by the chiral angle $\alpha$, 
with the rotated unit vectors $\overrightarrow{\bf{x}}$ and 
$\overrightarrow{\bf{y}}$ given by
\begin{align}
\left( \begin{matrix} \overrightarrow{\bf{x}} \\ \overrightarrow{\bf{y}} \end{matrix}\right) = \left( \begin{matrix} \cos(\alpha) & \sin(\alpha) \\ -\sin(\alpha) & \cos(\alpha) \end{matrix}\right) \left( \begin{matrix} \overrightarrow{\bf{x}}' \\ \overrightarrow{\bf{y}}' \end{matrix}\right)
\label{rotation matrix}
\end{align}
(see Fig.~\ref{chiral}). 
The arrangement of carbon atoms shows a new periodicity
along the direction of $\overrightarrow{\bf{y}}$, perpendicular to 
the chiral vector $\bf{L}$. The new period is the length $T$ of the 
translation vector ${\bf{T}}$, equal to
\begin{align}
T = \sqrt{3} a 	\frac{\sqrt{n^2_a + n^2_b - n_a n_b}}{\left| \text{GCD} \lbrace (n_a - 2 n_b), (2 n_a - n_b) \rbrace \right|},
\end{align}
where GCD$\lbrace n,m\rbrace$ is the greatest common divisor
between $n$ and $m$.
Vectors ${\bf{L}}$ and ${\bf{T}}$ define the CNT unit cell, which contains 
$N_{\text{A+B}}$ carbon atoms,
\begin{align}
N_{\text{A+B}} = \frac{4 (n^2_a + n^2_b - n_a n_b)}{\left| \text{GCD} \lbrace (n_a - 2 n_b), (2 n_a - n_b) \rbrace \right|}.
\end{align}
One has $\overrightarrow{\bf{x}} = {\bf{L}} / L$ 
and $\overrightarrow{\bf{y}} = {\bf{T}} / T$. 

The CNT is a cylinder of radius 
\begin{align}
R = \frac{L}{2 \pi} = \frac{a}{2 \pi} \sqrt{n^2_a + n^2_b - n_a n_b},
\end{align}
and axis parallel to ${\bf{T}}$. It is natural to use cylindrical coordinates, 
${\bf{r}} = \left( \rho, \theta, y\right)$, where $\rho \in [0, \infty)$ is the distance from the nanotube axis, 
$\theta \in [0, 2 \pi )$ is the azimuthal angle, and 
$y \in (-\infty, + \infty)$ is the axial coordinate. 
A vector lying in the original 2D graphene plane, 
${\bf{r}} = x \overrightarrow{\bf{x}} + y \overrightarrow{\bf{y}}$, 
is now described by coordinates $\rho = R$, $\theta = (x/R) \mod (2 \pi)$, 
and $y$. The origin of the reference frame is chosen such that 
atom ${\bf{R}}_{\text{B}}(0,0)$ in Eq.~\eqref{coordinates graphene} 
has coordinates $(\rho, \theta, y)  = (R, 0, 0)$. 
The cylindrical coordinates obtained from 
\eqref{coordinates graphene} are given by
\begin{align}
& \theta_{\text{A}}(n_1, n_2) =  \pi \! \left[ \frac{ n_a \left( 2 n_1 \! - \! n_2\right) - n_b \left( n_1 \! - \! 2 n_2 \! - \! 1 \right)}{n^2_a + n^2_b - n_a n_b} \! \! \! \! \! \mod 2 \right] , \nonumber \\
& y_{\text{A}}(n_1, n_2) = a \frac{\sqrt{3}}{2} 
\frac{n_a \left(n_2 + \frac{2}{3}\right) - n_b \left( n_1 + \frac{1}{3} \right)}{\sqrt{n^2_a + n^2_b - n_a n_b}}, \nonumber \\
& \theta_{\text{B}}(n_1, n_2) = \pi \! \left[ \frac{n_a \left( 2 n_1 - n_2\right) - n_b \left( n_1 - 2 n_2 \right)}{n^2_a + n^2_b - n_a n_b}  
\! \! \! \!\! \mod 2 \right] , \nonumber \\
& y_{\text{B}}(n_1, n_2) = a \frac{\sqrt{3}}{2} \frac{n_a n_2  - n_b n_1}{\sqrt{n^2_a + n^2_b - n_a n_b}}.
\label{coordinates}
\end{align} 

The set of equations \eqref{coordinates} maps the atomic positions of 
the original graphene plane into the locations of atoms on the CNT 
surface by letting $n_1$ and $n_2$ vary in $\mathbb{Z}$. 
A drawback is that there is an infinite number of atoms that are mapped 
into the same position on the CNT surface, i.e., those atoms with the same 
values of $y$ and $(x/R) \mod (2\pi)$. Since we will need to avoid 
multiple countings of atoms, it is more convenient to express atomic 
positions as a function of the two indexes $(n, j)$, 
unrelated to the original graphene geometry, defined as follows: 
$n$ fixes the axial coordinate and $j$ labels 
atoms lying on the same cross section of the CNT, given by $n$. 
The resulting expressions are
\begin{align}
& y_{\text{B}}(n) = \frac{\sqrt{3} a}{2 \sqrt{\nu_a^2 + \nu_b^2 - \nu_a \nu_b}} n, \nonumber \\
& \theta_{\text{B}}(n,j) = \left\{ \left[ n \theta_{\text{B}}(1) \right] \mod \left(\frac{2 \pi}{f_{ab}}\right) \right\}
+ j \frac{2 \pi}{f_{ab}}, \nonumber \\
& y_{\text{A}}(n) = y_{\text{B}}(n) + \Delta y_{\text{AB}} , \nonumber \\ 
& \theta_{\text{A}}(n,j) = \left\{ \left[ n \theta_{\text{B}}(1) + \Delta \theta_{\text{AB}}\right] \mod 
\left(\frac{2 \pi}{f_{ab}}\right) \right\} + j \frac{2 \pi}{f_{ab}},
\label{coordinates_nj}
\end{align}
with $n \in \mathbb{Z}$ for a tube of indefinite length, 
$j \in \lbrace 0, 1, \ldots, f_{ab}-1 \rbrace$, 
$f_{ab} = \text{GCD} \lbrace n_a , n_b  \rbrace $, with $n_a = f_{ab} \nu_a$ 
and $n_b = f_{ab} \nu_b$ so that integers 
$\nu_a$ and $\nu_b$ are coprime (if $n_b = 0$ then 
$f_{ab} = n_a$, $\nu_a = 1$, and $\nu_b = 0$), $\theta_{\text{B}}(1)$ 
is an angular offset depending on $n_a$ and $n_b$, whose expression 
is given in App.~\ref{a:coordinates} [Eq.~\eqref{theta_B(1)}], and 
\begin{align}
& \Delta y_{\text{AB}} = \frac{\sqrt{3} a}{2\sqrt{\nu_a^2 + \nu_b^2 - \nu_a \nu_b}} \frac{1}{3} \Big(2\nu_a - \nu_b\Big),
\nonumber \\
& \Delta \theta_{\text{AB}} = \frac{\nu_b \pi}{f_{ab} (\nu_a^2 + \nu_b^2 - \nu_a \nu_b)}.
\label{coordinates_offsets}
\end{align}
Equations \eqref{coordinates_nj}, \eqref{coordinates_offsets}, 
and \eqref{theta_B(1)} allow to uniquely determine the cylindrical 
coordinates of the atoms a nanotube of arbitrary chirality. 
Appendix \ref{a:coordinates} provides
the details of the derivation of Eqs.~\eqref{coordinates_nj} and 
\eqref{coordinates_offsets} starting from Eq.~\eqref{coordinates}.

\section{Many-body Hamiltonian}\label{s:MB}  

The many-body Hamiltonian, $\hat{H} = \hat{H}_{\text{SP}} + \hat{V}$, 
is the sum of two terms. The first one, $\hat{H}_{\text{SP}}$, 
is the single-particle Hamiltonian \eqref{H_SP} of a quantum dot
embedded in a semiconducting CNT, which includes kinetic energy, 
confinement potential, and spin-orbit coupling 
(see Appendix \ref{s:SP} for full details). 
The second one, $\hat{V}$, 
is the Coulomb interaction potential. 

We consider a gate-defined QD, whose  
confinement potential is a soft harmonic trap
of electrostatic origin:\cite{Kumar90,Reimann02}
\begin{align}
V_{\text{QD}}(y) = \frac{1}{2} m^* \omega_0^2 y^2,
\label{SQD}
\end{align}
with $m^*$ being the effective mass and $\omega_0$ the 
characteristic harmonic oscillator frequency.
The QD size in real space is given by the characteristic 
length $\ell_{\text{QD}} = \sqrt{\hbar / (m^* \omega_0)}$.
The Hamiltonian $\hat{H}_{\text{SP}}$ is 
written on the basis of the single-particle eigenstates as:
\begin{align}
\hat{H}_{\text{SP}} = \sum_{n} \sum_{\tau} \sum_{\sigma} \varepsilon_{n \tau \sigma} \hat{c}^{\dagger}_{n \tau \sigma} \hat{c}_{n \tau \sigma},
\label{H SP}
\end{align}
where $\hat{c}_{n \tau \sigma}$ destroys a fermion occupying the $n$th
harmonic-oscillator excited state with spin $\sigma$, isospin $\tau$,
and energy $\varepsilon_{n \tau \sigma}$
given by Eq.~\eqref{SO_disp}.

The Coulomb potential $\hat{V}$, which scatters different states
$\lbrace n, \tau, \sigma \rbrace$, is made of two 
terms,\cite{Secchi09,Secchi10} 
\begin{align}
\hat{V} = \hat{V}_{\text{FW}} + \hat{V}_{\text{BW}},
\end{align}
respectively for forward
\begin{align}
\hat{V}_{\text{FW}} = \frac{1}{2} \sum_{a b c d} \sum_{\tau \tau'} \sum_{\sigma \sigma'} V_{a, b; c, d}^{(\text{FW})} \hat{c}^{\dagger}_{a \tau \sigma} \hat{c}^{\dagger}_{b \tau' \sigma'} \hat{c}_{c \tau' \sigma'} \hat{c}_{d \tau \sigma}
\label{V_FW}
\end{align}
and backward scattering 
\begin{align}
\hat{V}_{\text{BW}} = \frac{1}{2} \sum_{a b c d} \sum_{\tau} \sum_{\sigma \sigma'} V_{a, b; c, d}^{(\text{BW})}(\tau) \hat{c}^{\dagger}_{a \tau \sigma} \hat{c}^{\dagger}_{b -\tau \sigma'} \hat{c}_{c \tau \sigma'} \hat{c}_{d -\tau \sigma}.
\label{V_BW}
\end{align}
Note that the FW term scatters different orbital states while 
conserving the individual isospins of the interacting electrons, 
whereas the BW term also exchanges the (opposite)
isospins of the interacting electrons. 
There is no BW term for electrons with like isospins. 
The quantities $V_{a, b; c, d}^{(\text{FW})}$ and 
$V_{a, b; c, d}^{(\text{BW})}(\tau)$, appearing respectively 
in Eqs.~\eqref{V_FW} and \eqref{V_BW}, are the two-body matrix elements of 
Coulomb interaction. We refer the reader to Ref.~\onlinecite{Secchi10}
for a detailed discussion of the FW term and focus on 
the BW term in the following.

\section{Backward scattering}\label{s:BW}  

This section is devoted to the analysis of the BW scattering term.
The starting point is our previous treatment of the BW potential as a contact
force, as reported in Ref.~\onlinecite{Secchi10}.
Here we extend our theory to include the effect of the CNT curvature.

\subsection{Backward scattering for the curved tube geometry}

We recall from Ref.~\onlinecite{Secchi10} [Eq.~(B5)] 
the generic expression of the two-body BW scattering
matrix element that appears in the operator \eqref{V_BW},
\begin{align}
& V^{(\text{BW})}_{a, b; c, d} (\tau)  \! = \! \frac{L_y^2}{4 N_c^2}	\ell_{\text{QD}}^{-2} 
\sum_{p, p'} \ee^{\ii \tau \phi_{p p'}} \! \! \sum_{\underline{R}_p} \! \sum_{\underline{R}'_{p'}} 
\ee^{\ii \tau \left( \underline{M}' - \underline{M} \right) 
 \cdot \left( \underline{R}_p - \underline{R}'_{p'} \right)} \nonumber \\
& \quad \times U\left(\left|\underline{R}_p - \underline{R}'_{p'}\right|\right)
F^*_{a}(y_p) F^*_{b}(y'_{p'}) F_{c}(y'_{p'}) F_{d}(y_p),
\label{BW_element}
\end{align}
where $p, p' \in \lbrace \text{A}, \text{B}\rbrace$ are the sublattice indexes, $\phi_{\text{AA}} = \phi_{\text{BB}} = 0$,
$\phi_{\text{AB}} = - \phi_{\text{BA}} = 2 \alpha + \frac{2 \pi}{3}$,
$\underline{M}$ and $\underline{M}' $ are the wave vectors of the 
conduction-band minima in the two valleys, 
$\underline{R}_p$ is the position of an atom of 
the $p$ sublattice, $U$ is the interaction potential, 
$N_c$ is the number of sublattice sites, $L_y$ is the CNT length,
and $F_n$ is the envelope function of the $n$th harmonic-oscillator state
(see also Appendix \ref{s:SP}). 
With respect to Eq.~(B5) of Ref.~\onlinecite{Secchi10}
here we have used cylindrical coordinates and included a minus
sign into phases $\phi_{\text{AB}}$.
Equation \eqref{BW_element} is derived exploiting the localization of
the $2p_z$ orbitals close to the atomic nuclei, 
whereas the envelope function $F_n(y)$ varies on the 
longer length scale $\ell_{\text{QD}}$, hence we assume
$\left| \phi_{2p_z}({\bf{r}} - {\bf{R}}_p) \right|^2 
\approx \delta({\bf{r}} - {\bf{R}}_p) \mathcal{V}_{\text{CNT}}$. 
Since this approximation washes out all effects 
related to the atomic orbitals, as an improvement 
we replace in Eq.~\eqref{BW_element}
the Coulomb potential with the Ohno potential,\cite{Ohno64,Ando06,Mayrhofer08} 
\begin{equation}
U({\bf{r}} - {\bf{r}}') = 
U_0 \left[ 1 + \epsilon^2 \left| {\bf{r}} - {\bf{r}}'\right|^2 
U_0^2 / e^4 \right]^{-1/2}, 
\end{equation}
which at short distances tends to the Hubbard-like value of 
the Coulomb repulsion between two electrons sitting on the $2p_z$ orbital, 
$U_0 \approx 15$ eV, whereas at long distances evolves into the 
screened Coulomb potential with static dielectric constant $\epsilon$. 

By making explicit the dependence of the atomic positions on the 
indexes $(n,j)$ as illustrated in Sec.~\ref{s:coordinates}, 
we write the interaction potential in cylindrical coordinates   
in the following symbolic form:
\begin{align}
& U\left[ \left|\underline{R}_p(n,j) - 
\underline{R}'_{p'}(n',j') \right| \right] \nonumber \\
& \equiv U \left\{ \left[ y_p(n) - y_{p'}(n')\right]^2, 
\sin^2\left[ \frac{\theta_p(n,j) - \theta_{p'}(n',j')}{2}\right]\right\}.
\label{U dependence}
\end{align}
After a lengthy calculation that is detailed in Appendix \ref{a:BW}, 
the matrix element \eqref{BW_element} is transformed into
\begin{align}
& V^{(\text{BW})}_{a, b; c, d}(\tau)  = \frac{\sqrt{3} a^2}{16 \pi R} \ell_{\text{QD}}^{-2} 
\int_{-\infty}^{+\infty} \text{d} y \sum_{n \in \mathbb{Z}}
\Big\{ \nonumber \\
& 2 f_{\tau}(n) F^*_{a}[y + y_{\text{B}}(n)] F^*_{b}(y) F_{c}(y) F_{d}[y + y_{\text{B}}(n)] \nonumber \\
& + g_{\tau}(n) F^*_{a}[y + y_{\text{A}}(n)] F^*_{b}(y) F_{c}(y) F_{d}[y + y_{\text{A}}(n)] \nonumber \\
& + g_{-\tau}(n) F^*_{a}(y) F^*_{b}[y + y_{\text{A}}(n)] 
F_{c}[y + y_{\text{A}}(n)] F_{d}(y) \Big\},
\label{BW matrix element}
\end{align}
where we have introduced two characteristic functions,
\begin{align}
& f_{\tau}(n) \equiv \ee^{\ii \tau \Delta M_k \cdot y_{\text{B}}(n)} \sum_{j = 0}^{f_{ab} - 1} \ee^{\ii \tau \Delta M_{\kappa} \cdot 
\theta_{\text{B}}(n,j)} U_{\text{B}}(n,j), \nonumber \\
& g_{\tau}(n) \equiv \ee^{\ii \tau \phi_{\text{AB}}} \ee^{\ii \tau \Delta M_k \cdot y_{\text{A}}(n)} \sum_{j = 0}^{f_{ab} - 1} \ee^{\ii \tau \Delta M_{\kappa} \cdot \theta_{\text{A}}(n,j)} U_{\text{A}}(n,j), 
\label{f and g}
\end{align}
with 
\begin{align}
(\Delta M_k, \Delta M_{\kappa}) \equiv (M'_k - M_k, M'_{\kappa} - M_{\kappa}),
\label{Delta M}
\end{align} 
and
\begin{align}
U_{p}(n,j) \equiv U\left\{ \left[ y_{p}(n) \right]^2, \sin^2\left[ \frac{\theta_{p}(n,j)}{2}\right] \right\}
\label{U_p}
\end{align}
for $p \in \lbrace \text{A, B}\rbrace$. 

The characteristic functions 
$f_{\tau}(n)$ and $g_{\tau}(n)$ determine the length scale and strength 
of BW interaction. In order to understand their physical meaning, we inspect 
the expression \eqref{BW matrix element} of the matrix element for BW 
scattering. The function $f_{\tau}(n)$ [$g_{\tau}(n)$] is weighted by 
the envelope functions of the interacting electrons, evaluated in positions 
along the CNT axis which are separated 
by $y_{\text{B}}(n)$ [$y_{\text{A}}(n)$]. 
Recalling the expressions \eqref{coordinates_nj} of the axial 
coordinates, we see that two coordinates differing by $y_{\text{B}}(n)$ 
belong to the same sublattice, while two coordinates differing 
by $y_{\text{A}}(n)$ belong to different sublattices. 
Therefore, the functions $f_{\tau}(n)$ and $g_{\tau}(n)$ measure 
the strength, 
respectively, of the intra- and inter-sublattice contributions to 
BW scattering. Moreover, since 
the distance between $y$ and $y + y_{\text{B(A)}}(n)$ is linear with 
$\left| n \right|$ and BW interaction is short-ranged, we expect 
$f_{\tau}(n)$ and 
$g_{\tau}(n)$ to vanish rapidly with increasing $\left| n \right|$,
as further discussed in subsection \ref{prop_f_and_g}.  
To gain a deeper insight into the properties of $f_{\tau}(n)$ and
$g_{\tau}(n)$,
it is convenient to work out the form of the BW scattering operator 
in first quantization, which is done in the following subsection.

\subsection{BW scattering potential in first quantization}

In this subsection we explicitly state the form of the BW scattering 
operator in the coordinate space representation. 

Let us introduce the isospinor $\varphi_{\tau}(t)$, 
which depends on the coordinate $t = \pm 1$ and is eigenstate 
of the isospin operator $\hat{\tau}$, 
with $\hat{\tau}(t) \varphi_{\tau}(t) = \tau \varphi_{\tau}(t)$. 
The electron has three coordinates: 
position along the axis $y$, spin $s$,
and isospin $t$, indicated as a whole by  
$\boldsymbol{z} \equiv (y, s, t)$. 
The wave function $\Psi_{n\sigma\tau}(\boldsymbol{z})$ is 
factorized as
\begin{align}
\Psi_{n \sigma \tau}(\boldsymbol{z}) = \left[ \ell_{\text{QD}}^{-1/2} F_n(y) \right] \otimes \chi_{\sigma}(s)
\otimes \varphi_{\tau}(t),
\label{SP with isospinor}
\end{align}
with the normalizations
\begin{align}
& \ell_{\text{QD}}^{-1} \int\text{d}y \; F^*_n(y) F_{n'}(y) = \delta_{n n'}, \nonumber \\
& \sum_s  \chi^*_{\sigma}(s) \chi_{\sigma'}(s) = \delta_{\sigma \sigma'}, \nonumber \\
& \sum_t  \varphi^*_{\tau}(t) \varphi_{\tau'}(t) = \delta_{\tau \tau'}.
\label{field norm}
\end{align}
As a straightforward generalization, the $N$-electron wave function,
$\Psi(\boldsymbol{z}_1, \boldsymbol{z}_2, \ldots, \boldsymbol{z}_N)$, 
depends on the set of orbital, 
$(y_1, y_2, \ldots, y_N)$, spin, $(s_1, s_2, \ldots, s_N)$, 
and isospin coordinates, $(t_1, t_2, \ldots, t_N)$. 

We look for the explicit expression of the BW scattering potential 
acting on $\boldsymbol{z}$ coordinates. It is easy to 
check that this must be a two-body potential 
of the form $\hat{V}_{\text{BW}}(y, t; y', t')$, 
acting on the orbital and isospin coordinates but not on spins. 
This is obtained by rewriting the second-quantized expression 
\eqref{V_BW} with the help of the isospinor formalism. 
In fact, the field annihilation operator is
\begin{align}
\hat{\Psi}(\boldsymbol{z}) \equiv \sum_{n} \sum_{\sigma} \sum_{\tau} \hat{\Psi}_{n \sigma \tau}(\boldsymbol{z}), 
\label{field operators}
\end{align}
with
\begin{align}
\hat{\Psi}_{n \sigma \tau}(\boldsymbol{z}) \equiv \Psi_{n \sigma \tau}(\boldsymbol{z}) \hat{c}_{n \sigma \tau}.
\end{align}
The BW term of the Hamiltonian is written 
in terms of the operator 
$\hat{V}_{\text{BW}}(\boldsymbol{z}; \boldsymbol{z}') \equiv 
\hat{V}_{\text{BW}}(y, t; y', t')$ as
\begin{align}
\hat{V}_{\text{BW}} = \frac{1}{2} \int\text{d}\boldsymbol{z} \int\text{d}\boldsymbol{z}'  
\hat{\Psi}^{\dagger}(\boldsymbol{z}) \hat{\Psi}^{\dagger}(\boldsymbol{z}') \hat{V}_{\text{BW}}(\boldsymbol{z}; \boldsymbol{z}') \hat{\Psi}(\boldsymbol{z}') \hat{\Psi}(\boldsymbol{z}),
\label{general 2body}
\end{align}
where $\int\text{d}\boldsymbol{z} \equiv \int\text{d}y \sum_s \sum_t$ 
and we mix operator symbols of first- and second-quantization.
After substituting the expansion \eqref{field operators} 
into \eqref{general 2body}, the result must be equal to \eqref{V_BW}. 
By further imposing the symmetry of the BW potential under 
coordinate permutation, 
$\hat{V}_{\text{BW}}(y, t; y', t') = \hat{V}_{\text{BW}}(y', t'; y, t)$, 
we obtain
\begin{align}
\hat{V}_{\text{BW}}(y, t; y', t') & = W_{\text{BW}}(y, y') \hat{\tau}^+(t) \hat{\tau}^-(t') \nonumber \\
& + W_{\text{BW}}(y', y) \hat{\tau}^-(t) \hat{\tau}^+(t') ,
\label{def W}
\end{align}
where $W_{\text{BW}}(y, y')$ is an operator acting on the orbital coordinates only, given by 
\begin{align}
& W_{\text{BW}}(y, y') =  \! \sum_{n \in \mathbb{Z}} \! \Big\{ \left[ f_{+1}(n) + f_{-1}(-n) \right] \delta\! \left[y' - y + y_{\text{B}}(n)\right]  \nonumber \\ 
& + g_{+1}(n) \delta\!\left[y' - y + y_{\text{A}}(n)\right] + g_{-1}(n) \delta\!\left[y' - y - y_{\text{A}}(n)\right] \Big\} \nonumber \\
& \times\frac{\sqrt{3} a^2}{16 \pi R}, 
\label{BW first Q}
\end{align}
and we have introduced the ladder operators of isospin:
\begin{align}
\begin{array}{ll}
\hat{\tau}^+(t) \varphi_{-1}(t) = \varphi_{+1}(t), & \hat{\tau}^+(t) \varphi_{+1}(t) = 0,\\
\hat{\tau}^-(t) \varphi_{-1}(t) = 0, & \hat{\tau}^-(t) \varphi_{+1}(t) = \varphi_{-1}(t).
\end{array}
\end{align}
These operators induce transitions between different conduction-band 
valleys as an effect of Coulomb interaction, exchanging the
crystal momentum of electrons.
We will see a similar effect with short-range disorder, which acts 
as a crystal momentum scatterer randomly placed in the CNT.

\subsection{Properties of functions $f$ and $g$}
\label{prop_f_and_g}

In this subsection we discuss the properties of the characteristic functions 
$f_{\tau}(n)$ and $g_{\tau}(n)$,  especially relevant
as their real parts determine the fine structure
of two-electron energy levels (cf.~Sec.~\ref{s:two}).
We consider
$\Re[f_{+1}] = \Re[f_{-1}] \equiv \Re[f]$ and 
$\Re[g_{+1}] = \Re[g_{-1}] \equiv \Re[g]$ for a few
representative tube geometries. Throughout the section we fix the 
dielectric constant as $\epsilon = 3.5$. 

\begin{figure}
\vspace{8mm}
\centerline{\epsfig{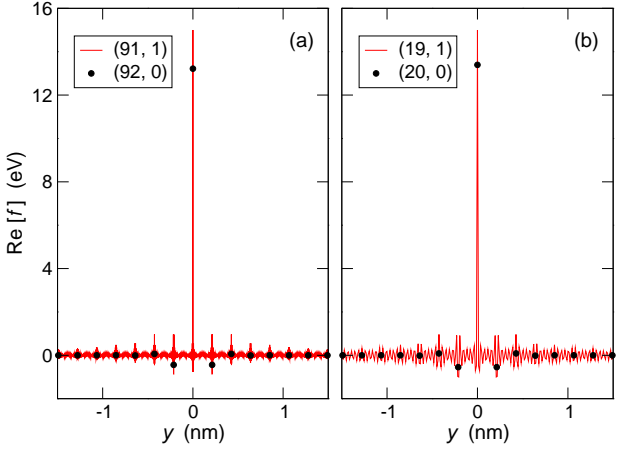}}
\caption{(Color online) $\Re[f]$ vs axial coordinate $y$, for CNTs 
with chiral angle close to $\alpha = 0$. 
(a) Tubes with a large radius: zigzag CNT with 
$(n_a, n_b) = (92, 0)$ and $R = 3.602$ nm
(black bullets) and chiral tube with $(n_a, n_b) = (91, 1)$ 
and $R = 3.543$ nm (red curve). 
(b) Tubes with a smaller radius: zigzag CNT with $(n_a, n_b) = (20, 0)$ 
and $R = 0.783$ nm (black bullets) and chiral tube with
$(n_a, n_b) = (19, 1)$ and $R = 0.725$ nm (red curve). 
Lines are guides to the eye. The parameters are 
$\epsilon = 3.5$ and $U_0 = 15$ eV.
\label{alpha_0_f}}
\end{figure}

\begin{figure}
\vspace{8mm}
\centerline{\epsfig{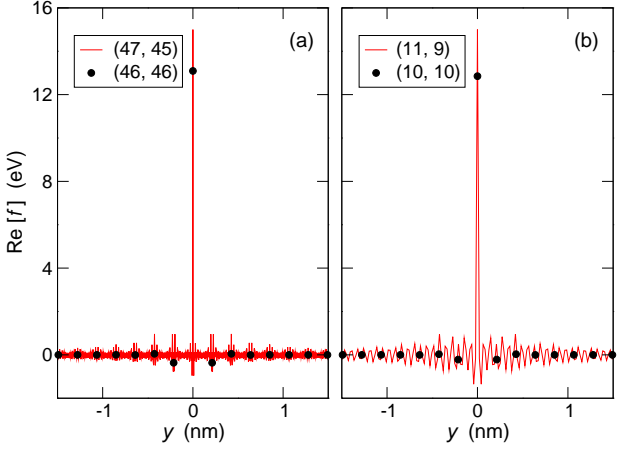}}
\caption{(Color online) $\Re[f]$ vs axial coordinate $y$, for CNTs 
with chiral angle close to $\alpha = \pi / 3$. 
(a) Tubes with a large radius: zigzag CNT with
$(n_a, n_b) = (46, 46)$ and $R = 1.801$ nm
(black bullets) and chiral tube with $(n_a, n_b) = (47, 45)$ 
and $R = 1.802$ nm (red curve).
(b) Tubes with a smaller radius: zigzag CNT with $(n_a, n_b) = (10, 10)$ 
and $R = 0.392$ nm (black bullets) and chiral tube with
$(n_a, n_b) = (11, 9)$ and $R = 0.397$ nm (red curve). 
Lines are guides to the eye. The parameters are
$\epsilon = 3.5$ and $U_0 = 15$ eV.
\label{alpha_60_f}}
\end{figure}

\begin{figure}
\vspace{8mm}
\centerline{\epsfig{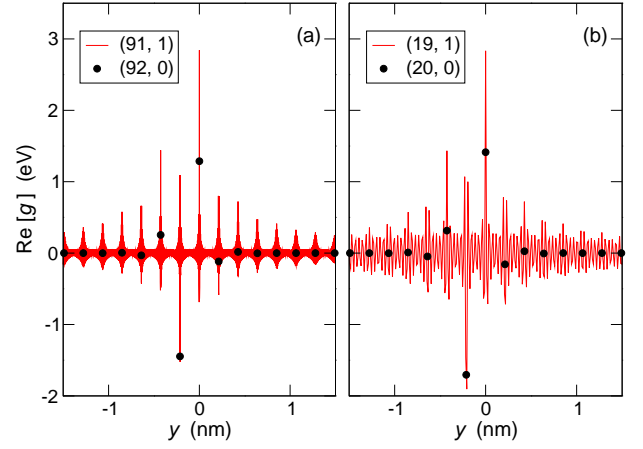}}
\caption{(Color online) $\Re[g]$ vs axial coordinate $y$, 
for CNTs with chiral angle close to $\alpha = 0$. The tubes
are the same as those studied in Fig.~\ref{alpha_0_f}.
\label{alpha_0_g}}
\end{figure}

\begin{figure}
\vspace{8mm}
\centerline{\epsfig{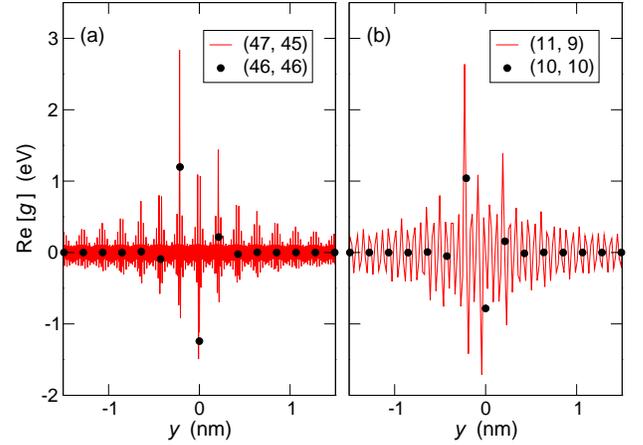}}
\caption{(Color online) $\Re[g]$ vs axial coordinate $y$, for CNTs with 
chiral angle close to $\alpha = \pi / 3$. 
The tubes are the same as those studied in Fig.~\ref{alpha_60_f}.
\label{alpha_60_g}}
\end{figure}

Equation \eqref{f and g} shows that $f$ ($g$) depends on the arrangement 
of the atoms in the B (A) sublattice. For semiconducting nanotubes, 
an examplar case is the zigzag configuration, with either 
$\alpha = 0$ ($n_b = 0$) or $\alpha = \pi/3$ ($n_a = n_b$). 
In this case $f$ as a function of the axial coordinate $y$
($n$) is even with respect 
to the origin (Figs.~\ref{alpha_0_f} and \ref{alpha_60_f}), whereas
the function $g$ does not have a definite symmetry,
as shown in Figs.~\ref{alpha_0_g} and \ref{alpha_60_g}. 
Indeed, the A sublattice is asymmetric with respect to $y = 0$:
for example, if $n_b = 0$, then for any $n_a$ we have $T = \sqrt{3} a$ 
and $\Delta y_{\text{AB}} = a / \sqrt{3} = T / 3$, so the A axial 
coordinates most close to $0$ are respectively $y_{\text{A}}(0) = T / 3$ and 
$y_{\text{A}}(-1) = -T / 6$. The zigzag configuration also maximizes 
the number of atoms on each allowed cross section and, conversely, 
minimizes the density of allowed $y$ coordinates along the tube axis. 
On the other hand, for generic chiral tubes there are 
more allowed axial coordinates with fewer atoms contributing to 
the circumferential cross section. Since in those cases even the 
arrangement of B atoms is not symmetric around $y = 0$, 
neither $f$ nor $g$ exhibit a well-defined symmetry.

Figure \ref{alpha_0_f}(a) shows $\Re[f]$ for the achiral ($\alpha = 0$)
zigzag tube $(n_a, n_b) = (92, 0)$ (black bullets) and chiral 
tube $(n_a, n_b) = (91, 1)$ (red curve)
obtained by applying a small twist 
$(-1, 1)$ to the zigzag one. 
The tube radius $R$ is approximately the same ($\approx 3.6$ nm) 
in both cases but the variation of the atom arrangement along the axis 
causes an appreciable variation of the profile of $f(n)$. 
In addition to the dominant maximum in the origin, 
$\Re[f]$ of the chiral tube exhibits many oscillations
on the length scale of $a/100$,
whereas the profile of the zigzag tube is smoother because
only a few axial coordinates are allowed.
Nevertheless, for such a large radius, the oscillations of $\Re[f]$ 
of the chiral tube are reminescent of those of the zigzag tube 
as the positions of the highest maxima overlap. 
For a smaller radius, the symmetry-breaking effect
of a $(-1, 1)$ twist of the zigzag tube is larger, as seen 
in Fig.~\ref{alpha_0_f}(b) for the tube $(n_a, n_b) = (20, 0)$. 
Apart from the central peak, the profiles of the chiral and zigzag
tubes now deviate more significantly than in Fig.~\ref{alpha_0_f}(a). 
Note that $\Re[f]$ with $\alpha = 0$ depends very weakly on the radius $R$ 
(i.e., $n_a$). 

Results for $\alpha = \pi / 3$ zigzag tubes with $n_a = n_b$ are shown
in Fig.~\ref{alpha_60_f} (black bullets)
for different radii ($R\approx$ 1.8 and 0.4 nm respectively in panels a and b), 
together with data
for tubes obtained by applying a $(1, -1)$ twist (red curves).
Although the CNTs with $\alpha=0$ and $\alpha = \pi / 3$ are
equivalent,
the functions plotted in Fig.~\ref{alpha_60_f} differ from those for 
$\alpha = 0$ because the arrangement of the atoms is shifted with respect to 
$y = 0$ in the two cases. This shows that $f$ depends strongly on the chiral 
angle. On the other hand, the comparison between chiral and achiral tubes
exhibits the same features as in Fig.~\ref{alpha_0_f}.

In Figs.~\ref{alpha_0_g} and \ref{alpha_60_g} we plot $\Re[g]$ for the 
eight nanotubes considered before. The $g$ function, which provides
the scattering between sublattices A and B, 
gives generically a weaker contribution to the BW Hamiltonian 
than the $f$ function,
which induces scattering within the same sublattice.
This may be seen by the difference between the maximum values of $\Re[f]$
(Figs.~\ref{alpha_0_f} and \ref{alpha_60_f})
and $\Re[g]$ (Figs.~\ref{alpha_0_g} and \ref{alpha_60_g}). 
Inspection of figures Figs.~\ref{alpha_0_g} and \ref{alpha_60_g}
also reveals that
the $g$ function for $\alpha = 0$ or $\alpha = \pi / 3$ depends very weakly 
on $R$ and that 
small distortions with respect to the zigzag configuration 
are sufficient to change significantly the profiles of $g(n)$, 
similarly to the features of function $f$.

The plots of $f$ and $g$ provide an insight into the features of 
BW scattering. Both $f(n)$ and $g(n)$ are significantly different from zero 
only close to $n = 0$, which confirms the short-range nature of BW 
interaction.\cite{Ando06} 
From Eq.~\eqref{BW first Q} it is clear that the dominant contribution 
comes from values of $f(n)$ close to $n = 0$, with $f(0)$ being 
of the order of the Hubbard parameter $U_0$. 
Function $g(n)$ has a nearly-zero average on a length scale of few 
nanometers (see Figs.~\ref{alpha_0_g} and \ref{alpha_60_g}), over which the 
QD envelope functions are not expected to vary appreciably, therefore 
its contribution is much smaller than that of $f$. 
Therefore, a first approximation is
$f(n) \approx U_0 \delta_{n 0}$ and $g(n) \approx 0$, which, 
applied to \eqref{BW first Q}, gives the approximated form of BW 
potential appearing in Eq.~\eqref{def W}:
\begin{align}
\hat{V}^{(0)}_{\text{BW}}(y, t; y', t') & \approx U_0 \frac{\sqrt{3} a^2}{8 \pi R} \delta(y - y') \nonumber \\
& \times \left[ \hat{\tau}^+(t) \hat{\tau}^-(t') + \hat{\tau}^-(t) \hat{\tau}^+(t') \right]. 
\end{align} 
This form reproduces the results of
Refs.~\onlinecite{Ando06, Secchi09, Secchi10}, showing that 
the BW scattering is expected to act significantly only on those 
many-body states in which electrons have a non-zero probability of being 
in contact. We study in detail the two-electron system in the next section. 

\section{Two electrons in a carbon-nanotube quantum dot}\label{s:two}  

In this section we recall from our previous 
studies\cite{Secchi09,Secchi10,Pecker13} the main features of 
the two-electron system in the absence of BW scattering. 
The effect of the BW interaction potential will be analyzed
in the next section.

The envelope-function Hamiltonian of two interacting electrons 
in a CNT QD is 
\begin{align}
\hat{H}(\boldsymbol{z}_1,\boldsymbol{z}_2) & = K \left( \partial^2_{y_1}, \partial^2_{y_2} \right) + 
V_{\text{QD}}\left(y_1, y_2\right) +  V_{\text{FW}}\left(\left|y_1-y_2\right|\right) \nonumber \\
& + \hat{V}_{\text{BW}}\left( y_1, t_1; y_2, t_2 \right) + \hat{H}_{\text{SO}}\left( s_1, t_1; s_2, t_2\right),
\label{full H}
\end{align}
where we have put the hat symbol only on the operators acting on spins and isospins. Here $K$ is the kinetic energy, 
$V_{\text{QD}}$ is the QD confinement potential, 
$V_{\text{FW}}$ is the FW scattering interaction (acting 
on orbital coordinates only), $\hat{V}_{\text{BW}}$ is the BW scattering 
interaction, and $\hat{H}_{\text{SO}}$ is the SO interaction, 
\begin{align}
\hat{H}_{\text{SO}}\left( s_1, s_2; t_1, t_2\right) = \nu \Delta_{\text{SO}}\frac{\gamma}{R} \left[ 
\hat{\sigma}(s_1) \hat{\tau}(t_1) + \hat{\sigma}(s_2) \hat{\tau}(t_2) \right].
\end{align}
We set
\begin{align}
&\hat{H} \equiv H_0 + \hat{H}', \nonumber \\
& H_0 \equiv K + V_{\text{QD}} + V_{\text{FW}} , \nonumber \\
&\hat{H}' \equiv \hat{V}_{\text{BW}} + \hat{H}_{\text{SO}},
\label{eq:Hpert}
\end{align}
assuming that $\hat{H}'$ can be treated as a small perturbation of $H_0$, 
as confirmed a posteriori by numerical evidence.\cite{Secchi09,Secchi10} 
The eigenvalue equation for $H_0$ is
\begin{align}
H_0 \Psi_{k, j}(\boldsymbol{z}_1,\boldsymbol{z}_2) = 
E_0(k) \Psi_{k, j}(\boldsymbol{z}_1,\boldsymbol{z}_2),
\end{align}
where the wave function may be factorized as
\begin{align}
\Psi_{k, j}(\boldsymbol{z}_1,\boldsymbol{z}_2) = 
\ell_{\text{QD}}^{-1} \psi_k(y_1, y_2) \otimes \xi_j(s_1, t_1; s_2, t_2) ,
\label{factorization}
\end{align}
with $\psi_k(y_1,y_2)$ being the orbital component and 
$\xi_j(s_1, t_1; s_2, t_2)$ the spin-valley component of the wave function. 
They are normalized as
\begin{align}
&\ell_{\text{QD}}^{-2} \int\!\! \text{d}y_1 \!\!\!\int\!\! \text{d}y_2 
\,\psi^*_k(y_1, y_2) \psi_{k'}(y_1, y_2) = \delta_{k k'}, \nonumber \\
& \sum_{s_1, s_2} \sum_{t_1, t_2} \xi_j^*(s_1, t_1; s_2, t_2) \xi_{j'}(s_1, t_1; s_2, t_2) \nonumber \\ 
&\quad = \delta_{j j'}.
\end{align}

The factorization \eqref{factorization} is always possible for
two electrons, hence both orbital and spin-valley wave functions 
have a definite symmetry under coordinate permutation 
while the total product $\Psi_{k, j}(\boldsymbol{z}_1,\boldsymbol{z}_2)$ 
is antisymmetric. It follows that the orbital and spin-valley parts 
are one even and the other one odd under particle exchange.
Since $H_0$ does not act on spin and isospin coordinates, 
the energy $E_0(k)$ depends only on the orbital component and is
possibly degenerate with respect to different spin-valley projections. 
The complete set of spin-valley functions for two electrons consists of 
six antisymmetric and ten symmetric components.\cite{Secchi09,Pecker13}
For example, the six antisymmetric spin-valley functions are obtained
by multiplying either a spin singlet times an isospin triplet or
a spin triplet times an isospin singlet (see also Tables 
\ref{t: S multiplet} and \ref{t: A multiplet}).
Therefore (in the absence of orbital degeneracy) $E_0(k)$ is either 
six-fold or ten-fold degenerate when $\psi_k$ is respectively even
or odd under coordinate exchange.\cite{Secchi09,Secchi10,Pecker13}  

We next discuss the features of the spectrum $E_0(k)$ in the case of 
harmonic confinement,
\begin{align}
V_{\text{QD}}(y_1, y_2) = 
\frac{1}{2} m^* \omega_0^2 \left( y_1^2 + y_2^2 \right).
\end{align}
Since the QD potential is quadratic and the interaction potential 
$V_{\text{FW}}$ depends on $\left| y_1 - y_2 \right|$ only, 
the canonical transformation to (normalized) center-of-mass (CM) and 
relative-motion (RM) coordinates
\begin{align}
y_{\text{CM}} = \frac{y_1 + y_2}{\sqrt{2}}, \quad y_{\text{RM}} = \frac{y_1 - y_2}{\sqrt{2}}
\end{align}
allows to separate the Hamiltonian $H_0$ into the sum of two terms,
\begin{align}
& H_0 \equiv H_{\text{CM}} + H_{\text{RM}}, \nonumber \\
& H_{\text{CM}} \equiv -\frac{\hbar^2}{2 m^*} \frac{\partial^2}{\partial y^2_{\text{CM}}} + \frac{1}{2} m^* \omega_0^2 y^2_{\text{CM}}, \nonumber \\
& H_{\text{RM}} \equiv -\frac{\hbar^2}{2 m^*} \frac{\partial^2}{\partial y^2_{\text{RM}}} + \frac{1}{2} m^* \omega_0^2 y^2_{\text{RM}} + V_{\text{FW}}\left( \sqrt{2} \left| y_{\text{RM}} \right| \right), 
\label{eq:HQD}
\end{align}
which depend separately on the coordinates $y_{\text{CM}}$ and 
$y_{\text{RM}}$. We may factorize the orbital wave function 
$\psi(y_1, y_2) \rightarrow \psi(y_{\text{CM}}, y_{\text{RM}})$ as 
\begin{align}
\psi_{n_{\text{CM}}, m}(y_{\text{CM}}, y_{\text{RM}}) \equiv F_{n_{\text{CM}}}(y_{\text{CM}}) \psi_m(y_{\text{RM}}),
\end{align}
where the CM wave function is determined by $H_{\text{CM}}$ and 
is an eigenstate of the harmonic oscillator, 
\begin{align}
& F_n(y) = u_n({\ell^{-1}_{\text{QD}} y}), \nonumber \\
& u_n(Y) = \left( \pi 2^n n! \right)^{-1/2} \text{e}^{-Y^2 / 2} \mathcal{H}_n(Y),
\label{SP orbitals}
\end{align}
with eigenvalue
\begin{align}
E\left(n_{\text{CM}}\right) = \left( n_{\text{CM}} + \frac{1}{2} \right) \hbar \omega_0,
\end{align}
for $n \in \lbrace 0, 1, 2, \ldots \rbrace$ ($\mathcal{H}_n(Y)$ is the Hermite
polynomial of order $n$). 
The problem associated with the RM wave function $\psi_m$ depends on the 
interaction and must be solved numerically. Since the CM wave function 
is symmetric under the interchange of $y_1$ and $y_2$, the symmetry of the 
total orbital wave function is the same as that of the RM wave function. 

\begin{figure}
\vspace{8mm}
\centerline{\epsfig{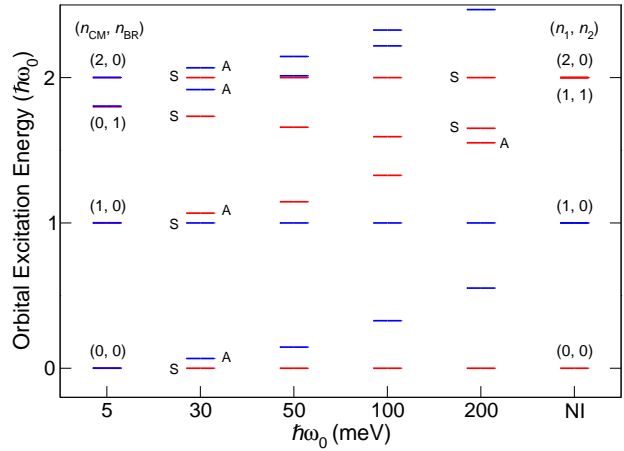}}
\caption{(Color online) Excitation energies of  two electrons, 
in units of $\hbar \omega_0$, vs $\hbar \omega_0$ (NI 
labels the limit $\hbar \omega_0 \rightarrow \infty$). 
Red (blue) levels point to states even (odd) under spatial reflection, 
$\lbrace y_1 \rightarrow - y_1, y_2 \rightarrow - y_2 \rbrace$,
whereas labels S (A) point to the (anti)symmetry of the 
wave function orbital component under particle exchange, 
$\lbrace y_1 \rightarrow  y_2, y_2 \rightarrow  y_1 \rbrace$.
The data are obtained from exact diagonalization, with
$\epsilon = 3.5$ and $R = 1$ nm. 
\label{E_0}}
\end{figure}

Figure \ref{E_0} shows the low-energy spectrum $E_0(k)$
associated to the Hamiltonian $H_0$ appearing in \eqref{eq:HQD},
obtained from exact 
diagonalization,\cite{Rontani06,Secchi09,Secchi10,Secchi12,Pecker13} 
as a function of the confinement strength $\hbar \omega_0$. 
The dielectric constant $\epsilon=3.5$ and the CNT radius $R=1$ nm
are typical values for Coulomb blockade experiments. 
The quantity on the vertical axis is the excitation energy, i.e., 
$E_0(k) - E_0(k=0)$, in units of $\hbar \omega_0$. 
This is ruled by the competition between
the energy scales respectively associated to the confinement potential, 
$\hbar \omega_0$, and FW Coulomb interaction. 
When $\hbar \omega_0$ is small the system is in the strongly-interacting 
Wigner molecule regime \cite{Secchi09,Secchi10,Secchi12,Pecker13} 
whereas when $\hbar \omega_0$ is large Coulomb interaction is negligible 
and the non-interacting (NI) picture holds. Below we consider in 
some detail the two limit regimes. 

\subsection{Non-interacting regime}

The NI regime is naturally described in the independent-particle framework. 
Orbital states are obtained as symmetrized or antisymmetrized products 
of single-particle orbitals $F_n(y)$ [Eq.~\eqref{SP orbitals}]. 
The quantum numbers $n_1$ and $n_2$ of the two orbitals occupied
identify the excited states whose wave functions are
\begin{align}
\psi_{\lbrace n_1, n_1 \rbrace} = F_{n_1}(y_1) F_{n_1}(y_2),
\label{n1 = n2}
\end{align}
if $n_1 = n_2$, and
\begin{align}
\psi_{\lbrace n_1, n_2 \rbrace}^{\pm} = \frac{1}{\sqrt{2}} \left[ F_{n_1}(y_1) F_{n_2}(y_2) \pm F_{n_2}(y_1) F_{n_1}(y_2)\right]
\label{NI functions}
\end{align} 
if $n_1 \ne n_2$, with excitation energies given by
\begin{align}
E^*(n_1, n_2) = \left( n_1 + n_2 \right) \hbar \omega_0.
\end{align} 
This picture, of course, may be recovered by alternatively 
using CM and RM coordinates. 
Expressions \eqref{n1 = n2} and \eqref{NI functions} show that states 
with $n_1 \ne n_2$ produce two orthogonal orbital wave functions 
$\psi_{\lbrace n_1, n_2 \rbrace}^{\pm}$, respectively symmetric $(+)$ 
and antisymmetric $(-)$ under particle exchange, whereas 
if $n_1 = n_2$ only the symmetric function is allowed (cf.~Fig.~\ref{E_0}). 
Therefore, a couple $(n_1, n_2)$ with $n_1 \ne n_2$ specifies a set of 
sixteen states, obtained by summing the ten-fold degenerate A states 
with the six-fold degenerate S states, while if $n_1 = n_2$ 
there are only six S states. Since energies depend on $(n_1 + n_2)$, 
we see that, e.g., the sets $(n_1, n_2) = (2, 0)$ and $(n_1, n_2) = (1, 1)$ 
are degenerate, with total degeneracy 
twenty-two. Note that
states belonging to a same shell have all the same orbital parity, 
equal to $(-1)^{n_1 + n_2}$,
as seen in the NI column of Fig.~\ref{E_0}.

\subsection{Wigner molecule}

The limit opposite to the NI regime is that of strong Coulomb repulsion. 
The low-energy states are then understood in terms of
a Wigner molecule made of electrons localized in space, 
arranged in the geometrical configuration that minimizes the 
Coulomb repulsion in the presence of the confinement 
potential.\cite{Secchi09,Secchi10,Secchi12,Pecker13} 
The competition between Coulomb potential and quantum confinement 
tunes the classical equilibrium positions of
the two electrons, $\pm \Lambda$, where
\begin{align}
\Lambda = \left( \frac{e^2}{4 \epsilon \omega_0^2 m^*} \right)^{1/3} 
\label{eq:lambda}
\end{align}
is located at the maximum of the particle density along the axis. 
Indeed, the density weight is concentrated in Gaussians centered at 
$\pm \Lambda$, whose finite widths originate
from the quantum fluctuations of the two electrons around their 
equilibrium positions. For a well-formed WM the Gaussian width 
is smaller than $\Lambda$, so the overlap between the localized
electrons is small. In this limit, one may 
safely expand the dominant long-range
part of Coulomb interaction, which goes like $\approx 1 / 
\left(\epsilon \left| y_1 - y_2 \right|\right)$, around the equilibrium 
positions up to quadratic order.\cite{Secchi10}

The approximated wave function is
\begin{align}
& \psi^{\pm}_{\lbrace n_{\text{CM}}, n_{\text{BR}} \rbrace} = F_{n_{\text{CM}}}(y_{\text{CM}}) \frac{3^{1/8}}{\sqrt{2} \mathcal{N}^{\pm}_{n_{\text{BR}}}} \nonumber \\
& \times \left\{ F_{n_{\text{BR}}}[3^{1/4} (y_{\text{RM}} - \sqrt{2} \Lambda)] \pm  F_{n_{\text{BR}}}[3^{1/4} (- y_{\text{RM}} - \sqrt{2} \Lambda)]
\right\},
\label{WM functions}
\end{align}
where the integer quantum number $n_{\text{BR}}$ counts the 
harmonic oscillation quanta of the antiphase 
normal mode known as breathing mode (BR). 
The BR characteristic frequency is $\sqrt{3} \omega_0$ and
the total WM excitation energies are 
\begin{align}
E^*(n_{\text{CM}}, n_{\text{BR}}) = \left( n_{\text{CM}} +  
n_{\text{BR}} \sqrt{3} \right) \hbar \omega_0 ,
\label{eq:WMWF}
\end{align}
as it may be checked in the $\hbar\omega_0=$ 5 meV column of Fig.~\ref{E_0}.
Symmetric $(+)$ and antisymmetric $(-)$ WM states 
with the same quantum numbers $(n_{\text{CM}}, n_{\text{BR}})$ are 
sixteen-fold degenerate, 
since the overlap between localized electrons is negligible.
This overlap enters Eq.~\eqref{WM functions} through the normalization 
constant $\mathcal{N}^{\pm}_{n_{\text{BR}}}$, which in turn depends on 
$n_{\text{BR}}$ and the symmetry $\pm$ of the RM wave function: for
strong correlations, $\mathcal{N}^{\pm}_{n_{\text{BR}}} \approx 1$. 
The formula \eqref{eq:WMWF} loses accuracy with increasing 
$n_{\text{BR}}$, since at higher energy the harmonic approximation
for the interaction potential breaks down.

In summary, in both the NI and WM regimes the allowed energy 
states are specifyed by two integer quantum numbers, 
respectively $(n_1, n_2)$ and $(n_{\text{CM}}, n_{\text{BR}})$. 
Figure \ref{E_0} shows the evolution of the energy spectrum 
between these two limits as $\hbar \omega_0$ is increased. 
The large dot with $\hbar \omega_0 = 5$ is in the WM regime, 
as recognized from the degeneracies of even and odd states and
their spacings, understood in terms of CM and BR 
excitations---see e.g.~the first BR excitation labelled 
$(n_{\text{CM}}, n_{\text{BR}}) = (0, 1)$ and the CM excitations 
$(1, 0)$ and $(2, 0)$. As $\hbar \omega_0$ is increased the degeneracy 
of even and odd states is lifted, so the spectrum is a sequence of 
multiplets of even (red color) or odd (blue)
spatial parity and symmetry (S or A)
under particle exchange. The spectrum then merges the NI regime, 
whose excitations $n \hbar \omega_0$ are equally spaced, each 
one with a well-defined spatial parity. 

\begin{table*}
\caption{Antisymmetric spin-valley wave functions for two electrons in CNTs.}
%\begin{ruledtabular}
\begin{tabular}{  | c | c | c | }
\hline
$\eta$   &     $\xi(s_1, t_1; s_2, t_2)$    &  $\xi^{\eta}_{\sigma, \tau}$         \\
\hline
-2   &    $\frac{1}{\sqrt{2}} \left[ \chi_{-1}(s_1) \chi_{+1}(s_2) \cdot \varphi_{+1}(t_1) \varphi_{-1}(t_2) - \chi_{+1}(s_1) \chi_{-1}(s_2) \cdot \varphi_{-1}(t_1) \varphi_{+1}(t_2) \right] $  &               $\xi^{-2}_{0, 0}$  \\    
\hline
0   &    $\chi_{-1}(s_1) \chi_{-1}(s_2) \cdot \frac{1}{\sqrt{2}} \left[ \varphi_{+1}(t_1) \varphi_{-1}(t_2) - \varphi_{-1}(t_1) \varphi_{+1}(t_2)     \right]  $  & 
             $\xi^{0}_{-2, 0}$ \\
    &    $\frac{1}{\sqrt{2}} \left[ \chi_{+1}(s_1) \chi_{-1}(s_2) - \chi_{-1}(s_1) \chi_{+1}(s_2)  \right] \cdot \varphi_{-1}(t_1) \varphi_{-1}(t_2) $  & 
             $\xi^{0}_{0, -2}$ \\
    &    $\frac{1}{\sqrt{2}} \left[ \chi_{+1}(s_1) \chi_{-1}(s_2) - \chi_{-1}(s_1) \chi_{+1}(s_2)  \right] \cdot \varphi_{+1}(t_1) \varphi_{+1}(t_2) $  & 
             $\xi^{0}_{0, +2}$ \\
    &    $\chi_{+1}(s_1) \chi_{+1}(s_2) \cdot \frac{1}{\sqrt{2}} \left[ \varphi_{+1}(t_1) \varphi_{-1}(t_2) - \varphi_{-1}(t_1) \varphi_{+1}(t_2) \right] $  & 
             $\xi^{0}_{+2, 0}$ \\
\hline
+2   &    $\frac{1}{\sqrt{2}} \left[ \chi_{-1}(s_1) \chi_{+1}(s_2) \cdot \varphi_{-1}(t_1) \varphi_{+1}(t_2) - \chi_{+1}(s_1) \chi_{-1}(s_2) \cdot \varphi_{+1}(t_1) \varphi_{-1}(t_2) \right] $  & 
             $\xi^{+2}_{0, 0}$ \\
\hline
\end{tabular}
%\end{ruledtabular}
\label{t: S multiplet}
\end{table*}

\begin{table*}
\caption{Symmetric spin-valley wave functions for two electrons in CNTs.}
%\begin{ruledtabular}
\begin{tabular}{ | c | c | c | }
\hline
$\eta$   &     $\zeta(s_1, t_1; s_2, t_2)$  & $\zeta^{\eta}_{\sigma, \tau}$   \\
\hline
-2   &    $\chi_{-1}(s_1) \chi_{-1}(s_2) \cdot \varphi_{+1}(t_1) \varphi_{+1}(t_2)$  &  $\zeta^{-2}_{-2, +2}$ \\
     &    $\frac{1}{\sqrt{2}} \left[ \chi_{-1}(s_1) \chi_{+1}(s_2) \cdot \varphi_{+1}(t_1) \varphi_{-1}(t_2) + \chi_{+1}(s_1) \chi_{-1}(s_2) \cdot \varphi_{-1}(t_1) \varphi_{+1}(t_2) \right]$ & 
            $\zeta^{-2}_{0, 0}$  \\
     &    $\chi_{+1}(s_1) \chi_{+1}(s_2) \cdot \varphi_{-1}(t_1) \varphi_{-1}(t_2)$  &  $\zeta^{-2}_{+2, -2}$ \\     
\hline
0   &    $\chi_{-1}(s_1) \chi_{-1}(s_2) \cdot \frac{1}{\sqrt{2}} \left[ \varphi_{+1}(t_1) \varphi_{-1}(t_2) + \varphi_{-1}(t_1) \varphi_{+1}(t_2) \right]$  & $\zeta^0_{-2, 0}$ \\
    &    $\frac{1}{\sqrt{2}} \left[ \chi_{+1}(s_1) \chi_{-1}(s_2) + \chi_{-1}(s_1) \chi_{+1}(s_2)  \right] \cdot \varphi_{-1}(t_1) \varphi_{-1}(t_2)$  & $\zeta^0_{0, -2}$ \\
    &    $\frac{1}{\sqrt{2}} \left[ \chi_{+1}(s_1) \chi_{-1}(s_2) + \chi_{-1}(s_1) \chi_{+1}(s_2)  \right] \cdot \varphi_{+1}(t_1) \varphi_{+1}(t_2)$  & $\zeta^0_{0, +2}$ \\
    &    $\chi_{+1}(s_1) \chi_{+1}(s_2) \cdot \frac{1}{\sqrt{2}} \left[ \varphi_{+1}(t_1) \varphi_{-1}(t_2) + \varphi_{-1}(t_1) \varphi_{+1}(t_2) \right]$  & $\zeta^0_{+2, 0}$ \\
\hline
+2   &    $\chi_{-1}(s_1) \chi_{-1}(s_2) \cdot \varphi_{-1}(t_1) \varphi_{-1}(t_2)$ &   $\zeta^{+2}_{-2, -2}$ \\
     &    $\frac{1}{\sqrt{2}} \left[ \chi_{-1}(s_1) \chi_{+1}(s_2) \cdot \varphi_{-1}(t_1) \varphi_{+1}(t_2) + \chi_{+1}(s_1) \chi_{-1}(s_2) \cdot \varphi_{+1}(t_1) \varphi_{-1}(t_2) \right]$ & 
            $\zeta^{+2}_{0, 0}$  \\
     &    $\chi_{+1}(s_1) \chi_{+1}(s_2) \cdot \varphi_{+1}(t_1) \varphi_{+1}(t_2)$ &   $\zeta^{+2}_{+2, +2}$ \\
\hline
\end{tabular}
%\end{ruledtabular}
\label{t: A multiplet}
\end{table*}

\section{Backward scattering in the two-electron system}\label{s:two_plus_BW}

So far we discussed orbital excitations $E_0(k)$ of two electrons 
that are highly degenerate in
the spin-valley sector. The perturbation $\hat{H}'$, as defined
in Eq.~\eqref{eq:Hpert}, includes SO and BW interactions that
act on the spin-valley component of the wave function,
 splitting the energy levels 
within each orbital multiplet. 
Assuming that the energy spacings $E_0(k') - E_0(k)$ 
between orbital multiplets for 
any $k' \ne k$ are large with respect to the perturbation strength, 
in this section
we apply first-order degenerate perturbation theory to derive the 
multiplet fine structure. 

Table \ref{t: S multiplet} (\ref{t: A multiplet}) lists 
the six antisymmetric  (ten symmetric) spin-valley wave functions 
$\xi(s_1,t_1;s_2,t_2)$ [$\zeta(s_1,t_1;s_2,t_2)$].  
SO interaction splits the levels according to the
total helicity $\eta \in \lbrace -2, 0, +2 \rbrace$,
defined as 
$\eta = \sigma_1\tau_1 + \sigma_2\tau_2$, shown in the left column
of both Tables. 
We consider the two-electron ground state, whose orbital wave function
is symmetric (S), diagonalizing $\hat{H}'$ on the basis of the six spin-valley 
antisymmetric 
wave functions $\xi(s_1,t_1;s_2,t_2)$ of Table \ref{t: S multiplet}. 
It is convenient to represent $\xi(1,2)$ vectorially in the following.
Introducing the column vectors
\begin{align}
& \chi_{+1}(s_j) \rightarrow \left( \begin{matrix} 1 \\ 0 \end{matrix} \right)_j \! \equiv \chi_{+1}(j), \quad \chi_{-1}(s_j) \rightarrow \left( \begin{matrix} 0 \\ 1 \end{matrix} \right)_j \! \equiv \chi_{-1}(j), \nonumber \\
& \varphi_{+1}(t_j) \rightarrow \left[ \begin{matrix} 1 \\ 0 \end{matrix} \right]_j \equiv  \varphi_{+1}(j), \quad \varphi_{-1}(t_j) \rightarrow \left[ \begin{matrix} 0 \\ 1 \end{matrix} \right]_j \equiv \varphi_{-1}(j),
\end{align}
we compactly write their product as 
\begin{align}
& \chi_{\sigma_1}(s_1) \chi_{\sigma_2}(s_2) \varphi_{\tau_1}(t_1) \varphi_{\tau_2}(t_2) \nonumber \\
& \rightarrow \chi_{\sigma_1}(1) \otimes \chi_{\sigma_2}(2) \otimes \varphi_{\tau_1}(1) \otimes \varphi_{\tau_2}(2) \equiv \xi(1,2),
\end{align}
with $\xi_i^{\dagger}(1,2) \cdot \xi_j(1,2) = \delta_{i,j}$. Consistently, the isospin operators assume a matrix form:
\begin{align}
& \hat{\tau}(t_j) \rightarrow   \left[ \begin{matrix} 1 & 0 \\ 0 & -1 \end{matrix} \right]_j \equiv \hat{\tau}_j, \nonumber \\
& \hat{\tau}^+(t_j) \rightarrow \left[ \begin{matrix} 0 & 1 \\ 0 & 0  \end{matrix} \right]_j \equiv \hat{\tau}_j^+, \nonumber \\
& \hat{\tau}^-(t_j) \rightarrow \left[ \begin{matrix} 0 & 0 \\ 1 & 0  \end{matrix} \right]_j \equiv \hat{\tau}_j^-.
\end{align}

The perturbation matrix elements may then be written as
\begin{align}
H'_{i j}(k) & = \ell_{\text{QD}}^{-2} \int\text{d}y_1 \int\text{d}y_2  \; \xi_i^{\dagger}(1,2) \otimes \psi_k^*(y_1, y_2) \nonumber \\ 
& \times \left[ \hat{H}_{\text{SO}}(1,2) + \hat{V}_{\text{BW}}(1,2) \right] \psi_{k}(y_1, y_2) \otimes \xi_j(1,2)
\nonumber \\
& = \xi_i^{\dagger}(1,2) \cdot \hat{H}_{\text{SO}}(1,2) \cdot \xi_j(1,2) \nonumber \\
& + \xi_i^{\dagger}(1,2) \cdot \left( \hat{\tau}^+_1 \hat{\tau}^-_2 + \hat{\tau}^-_1 \hat{\tau}^+_2 \right) \cdot 
\xi_j(1,2) \nonumber \\
& \times \ell_{\text{QD}}^{-2} \int\text{d}y_1 \int\text{d}y_2  \; \left| \psi_{k}(y_1, y_2) \right|^2  W_{\text{BW}}(y_1, y_2)  ,
\label{pert 2}
\end{align} 
where we have used the symmetry of $\left| \psi_{k}(y_1, y_2) \right|^2$ 
under the permutation of $y_1$ and $y_2$ ($k=0$ for the ground state). 
Substituting Eq.~\eqref{BW first Q} into Eq.~\eqref{pert 2}, 
and noting that 
$f_{+1}(n) = f^*_{-1}(n)$, $g_{+1}(n) = g^*_{-1}(n)$, one obtains
\begin{align}
H'_{i j}(k)  = & \xi_i^{\dagger}(1,2) \cdot \hat{H}_{\text{SO}}(1,2) \cdot \xi_j(1,2) \nonumber \\
& + \Delta E_{\text{BW}}(k) \, \xi_i^{\dagger}(1,2) \cdot \left[ \hat{\tau}^+_1 \hat{\tau}^-_2 + \hat{\tau}^-_1 \hat{\tau}^+_2 \right] \cdot \xi_j(1,2) .
\label{second to last}
\end{align}
The key quantity $\Delta E_{\text{BW}}(k)$ appearing in
\eqref{second to last} is defined as
\begin{align}
\Delta E_{\text{BW}}(k) & \equiv \ell_{\text{QD}}^{-2} \sum_{n \in \mathbb{Z}}   \; \Big\{ 
\Re[f(n)] \int \Big| \psi_{k}[y, y + y_{\text{B}}(n)] \Big|^2 \text{d}y
\nonumber \\
& + \Re[g(n)] \int \Big| \psi_{k}[y, y + y_{\text{A}}(n)] \Big|^2 \text{d}y  \Big\} \frac{\sqrt{3} a^2}{8 \pi R}, 
\label{Delta E_BW}
\end{align}
where
\begin{align}
P_k (x) \equiv \int \Big| \psi_{k}(y, y + x) \Big|^2 \text{d}y 
\label{P_k}
\end{align}
is the pair correlation function 
associated with the orbital wave function $\psi_k(y_1, y_2)$. 

Since functions $f(n)$ and $g(n)$ are peaked close to $n = 0$ and 
decrease fast with increasing $\left| n \right|$ 
(cf.~Sec.~\ref{prop_f_and_g}), the leading contribution to 
$\Delta E_{\text{BW}}(k)$ is given by $P_k(x)$ for $x \approx 0$. 
This is consistent with the fact that BW interaction is short-range, 
as $P_k(0)$ is the probability for the two electrons to be in the 
same position along the axis. 
For this very reason BW scattering is inefficient in the 
excited A multiplet, 
as $P_{k = \text{A}}(0) = 0$. Therefore, 
we shall focus on the S low-energy multiplet only.

\begin{figure}
%\vspace{4mm}
\centerline{\epsfig{file=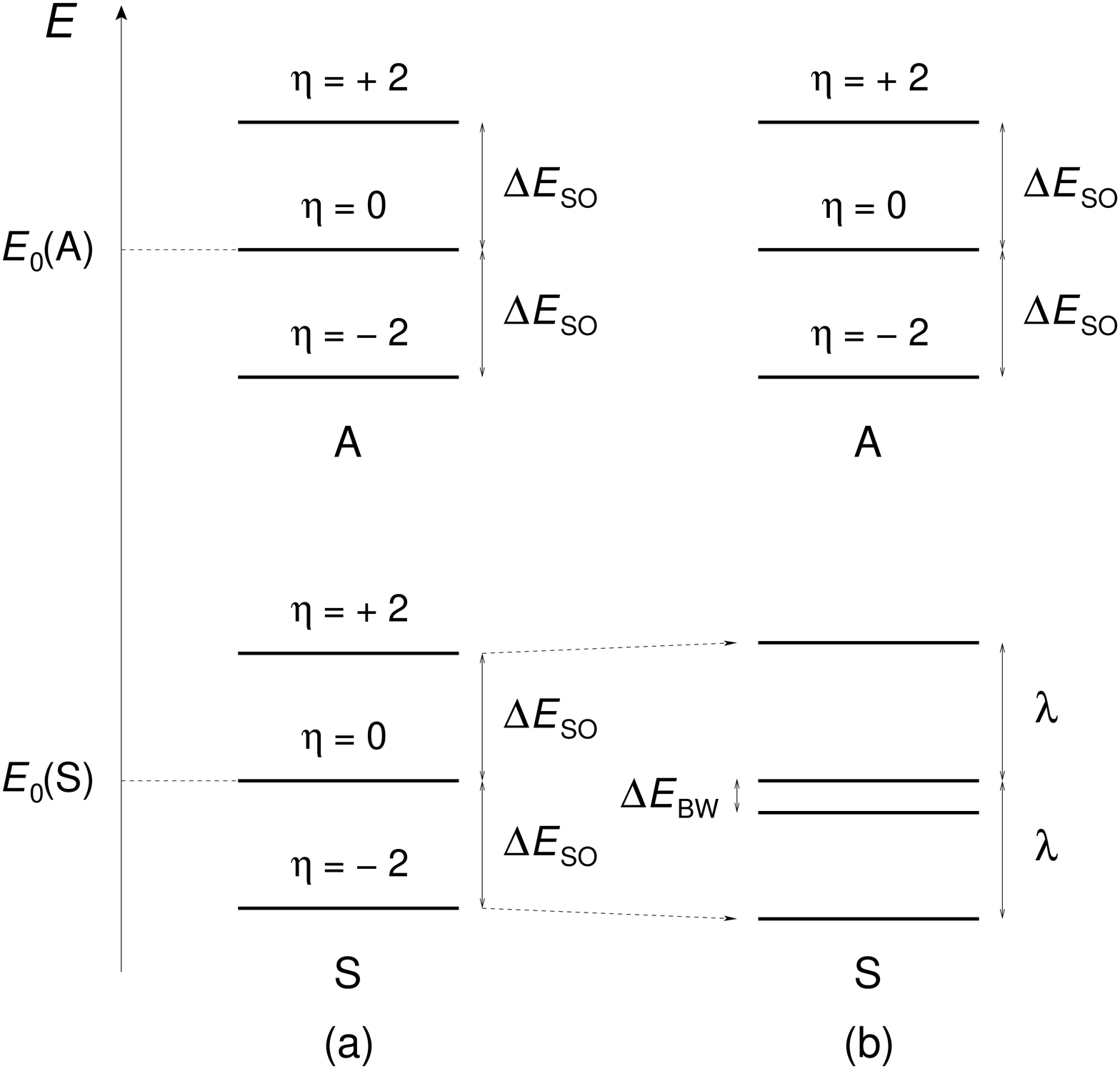,width=3in,angle=0}}
\caption{Fine structure of the two lowest orbital multiplets of two 
electrons. $E_0(\text{S})$ and $E_0(\text{A})$ are the energies 
of symmetric (S) and antisymmetric (A) orbital levels, respectively, 
with $E_0(\text{A}) - E_0(\text{S}) > 2 \Delta E_{\text{SO}}$. 
(a) Only SO coupling is taken into account, hence
the total helicity $\eta$ is a good quantum number and each multiplet 
is split into three equally spaced energy levels. 
In order of increasing energy, the level degeneracy 
of the S (A) multiplet is 1, 4, 1 (3, 4, 3).
(b) Both SO coupling and BW scattering are considered, 
hence S states with $\eta = 0$ are split by $\Delta E_{\text{BW}}$, 
while the highest and lowest S levels have now energies 
$E_0(\text{S}) \pm \lambda$ [see Eq.~\eqref{lambda}] with 
mixed helicities. BW interaction hardly affects the A multiplet, 
unaltered with respect to the case (a).
In order of increasing energy, the level degeneracy
of the S (A) multiplet is 1, 2, 2, 1 (3, 4, 3).
\label{energy levels All}}
\end{figure}

It is useful to make the notation more compact,
labelling the spin-valley functions as 
$\xi^{\eta}_{\sigma , \tau}$,
according to the right column of Table \ref{t: S multiplet}. 
This allows to link the levels to the corresponding
eigenstates of $H_0 + \hat{H}_{\text{SO}}$, identified by 
the quantum numbers $\eta = \eta_1 + \eta_2$, 
$\sigma = \sigma_1 + \sigma_2$, 
and $\tau = \tau_1 + \tau_2$, as shown in column (a) of
Fig.~\ref{energy levels All}.
It is clear from the structure of Eq.~\eqref{second to last} that
BW scattering acts on states with $\tau = 0$ only, 
whereas SO coupling acts on states with $\eta \ne 0$. 
Note that the total spin projection $\sigma = \sigma_1 + \sigma_2$ 
remains a good quantum number. 

Among the six states of the $\psi_{k = \text{S}}$ multiplet:
\begin{enumerate}
\item states $\xi^{0}_{0 , +2}$ and 
$\xi^{0}_{0 , -2}$ are not affected by $\hat{H}'$;
\item states $\xi^{0}_{+2 , 0}$ and 
$\xi^{0}_{-2 , 0}$ are affected only by 
$\hat{V}_{\text{BW}}$ but not mixed;
\item states $\xi^{+2}_{0 , 0}$ and 
$\xi^{-2}_{0 , 0}$ are affected by 
$\hat{H}_{\text{SO}}$ 
and mixed by $\hat{V}_{\text{BW}}$. 
\end{enumerate}

Focusing on those states affected by BW interaction, 
the two states $\xi^{0}_{+2 , 0}$ and 
$\xi^{0}_{-2 , 0}$ untouched by SO coupling, 
\begin{align}
\xi^{0}_{\pm 2 , 0}(1,2) = & \frac{1}{\sqrt{2}} \left[ \varphi_{+1}(1) \varphi_{-1}(2) - \varphi_{-1}(1) \varphi_{+1}(2) \right] \nonumber \\
& \otimes \chi_{\pm 1}(1) \chi_{\pm 1}(2),
\end{align}
remain unchanged in their form and are shifted in energy 
by the expectation value of $\hat{V}_{\text{BW}}$.  
Since
\begin{align}
\left( \hat{\tau}^+_1 \hat{\tau}^-_2 + \hat{\tau}^-_1 \hat{\tau}^+_2 \right) \xi^{0}_{\pm 2 , 0}(1,2) 
= - \xi^{0}_{\pm 2 , 0}(1,2) ,
\end{align} 
we obtain 
\begin{align}
\left< \hat{H}' \right>_{\sigma = \pm 2}^{\eta = 0} = - \Delta E_{\text{BW}} .
\end{align}
Therefore, the two states are degenerate with a total energy equal 
to $E_0 - \Delta E_{\text{BW}}$.

The other two states with $\eta \ne 0$, 
\begin{align}
\xi^{\pm 2}_{0 , 0}(1,2) = \frac{1}{\sqrt{2}} \sum_{\alpha = -1, +1} \alpha \chi_{\alpha}(1) 
\chi_{- \alpha}(2) \otimes \varphi_{\pm \alpha}(1) \varphi_{\mp \alpha}(2),
\end{align}
are mixed by $\hat{V}_{\text{BW}}$. 
Since 
\begin{align}
\left( \hat{\tau}^+_1 \hat{\tau}^-_2 + \hat{\tau}^-_1 \hat{\tau}^+_2 \right) \xi^{\pm 2}_{0 , 0}(1,2) = 
\xi^{\mp 2}_{0 , 0}(1,2) 
\end{align}
the mixing matrix is given by
\begin{align}
\mathbb{H}'_{\eta \ne 0} = \left( \begin{matrix} \nu \Delta E_{\text{SO}} & \Delta E_{\text{BW}} \\ 
                                    \Delta E_{\text{BW}} & - \nu \Delta E_{\text{SO}} \end{matrix}\right) ,
\label{mixing matrix}
\end{align}
with $\Delta E_{\text{SO}} = 2 \Delta_{\text{SO}} \gamma / R$. 
Diagonalization of \eqref{mixing matrix} yields the 
eigenvalues 
\begin{align}
\pm \sqrt{(\Delta E_{\text{SO}})^2 + (\Delta E_{\text{BW}})^2} \equiv \pm \lambda, 
\label{lambda}
\end{align}
whose eigenstates are
\begin{align}
& \xi_{0, 0}^{(+)}(1,2) = \frac{\Delta E_{\text{BW}} \, \xi^{+2}_{0 , 0}(1,2) + \left( \lambda - \nu \Delta E_{\text{SO}} \right) \, \xi^{-2}_{0 , 0}(1,2)}{\sqrt{2 \lambda \left(\lambda -\nu \Delta E_{\text{SO}} \right) }} \nonumber \\
& \xi_{0, 0}^{(-)}(1,2) = \frac{ \left( \lambda - \nu \Delta E_{\text{SO}} \right) \, \xi^{+2}_{0 , 0}(1,2) - \Delta E_{\text{BW}} \, \xi^{-2}_{0 , 0}(1,2)}{\sqrt{2 \lambda \left(\lambda -\nu \Delta E_{\text{SO}} \right) } } .
\end{align}
The above results for the fine structure of the lowest S multiplet 
are illustrated
in Fig.~\ref{energy levels All} in the presence of SO coupling only (a) 
as well as in combination with BW interaction (b). 

We have evaluated the quantity $\Delta E_{\text{BW}}$ by first 
performing exact diagonalization 
calculations\cite{Secchi09,Secchi10,Secchi12,Pecker13} in order to 
find the eigenstates of the two-electron Hamiltonian 
$H_0 + \hat{H}_{\text{SO}}$
[cf.~\eqref{eq:Hpert}],
from which we obtain the pair correlation functions 
$P_k(x)$, as defined in Eq.~\eqref{P_k}, and then apply the 
formula \eqref{Delta E_BW}. 
To evaluate the impact of the BW term,
we have considered realistic values of the confinement potential, 
$\hbar \omega_0 \in \lbrace 5, 10, 15 \rbrace$ meV, 
dielectric constant $\epsilon \in \lbrace 2.4, 3.5, 4.5 \rbrace$, 
and radius $R \in \lbrace 1.018, 2.036, 2.976 \rbrace$ nm,
combining them in all possible ways. 
For each value of radius we have found all chiral numbers 
$(n_a, n_b)$ corresponding to tubes with $\alpha \in [0, \pi/3]$ with
a tolerance of $0.01$ nm on $R$. In this manner we have obtained 
respectively 10, 18 and 18 CNTs for $R =$ 1.018, 2.036, and 2.976 nm. 

Some significant results are reported in
Table \ref{t: Delta E_BW example}, 
showing that $\Delta E_{\text{BW}}$ is nearly insensitive to the 
chiral angle $\alpha$ and depends only on the confinement potential
$\hbar\omega_0$. Indeed,
$\Delta E_{\text{BW}}$ is of the order of some $\mu$eV up to a few 
tens of $\mu$eV, and the variation with $\alpha$ is of the order of a 
few tenths of $\mu$eV at most over the whole range $\alpha \in [0, \pi/3]$. 
More generally, $\Delta E_{\text{BW}}$ depends significantly
on the radius $R$ and $\epsilon$ as well as on $\hbar\omega_0$ but
very weakly on $\alpha$ (data not shown)---likely  
an effect of the slowly-varying confinement potential.

An overview of our systematic analysis is presented in 
Table \ref{t: Delta E_BW all}, where for each set of 
parameters $(R, \hbar \omega_0, \epsilon)$ we report the value 
of $\Delta E_{\text{BW}}$ (in $\mu$eV), averaged over the different 
chiral angles. It turns out that $\Delta E_{\text{BW}}$ is in a range 
between a few $\mu$eV up to tens of $\mu$eV. As a reference,
measured values of $\Delta E_{\text{SO}}$ are of the order of some 
hundreds of $\mu$eV 
(e.g., $\Delta E_{\text{SO}} \approx 370 \, \mu$eV in 
Ref.~\onlinecite{Kuemmeth08}), so the predicted value of
$\Delta E_{\text{BW}}$ is within one order of magnitude. 

Whereas $\Delta E_{\text{SO}}$ is inversely proportional to $R$ 
we find that $\Delta E_{\text{BW}}$ increases with $R$. 
This is due to the fact that the long-ranged FW interaction is
reduced,\cite{Secchi09,Wunsch09,Secchi10} thereby favoring electrons 
to be closer one to the other, which in turn makes the 
short-ranged BW interaction more effective. The increase of the 
dielectric constant $\epsilon$ and/or confinement energy
$\hbar\omega_0$ produce a similar effect, as seen 
in Table \ref{t: Delta E_BW all}.

\begin{table*}
\caption{Selected values of $\Delta E_{\text{BW}}$  as a function of chirality
$\alpha$ and confinement energy $\hbar\omega_0$.
The dielectric constant is $\epsilon = 3.5$ and 
the radius is $R = (2.036 \pm 0.005)$ nm.}
%\begin{ruledtabular}
\begin{tabular}{ | c | c | c | c | c |}
\hline
$(n_a, n_b)$    &     $\alpha$		& $\Delta E_{\text{BW}} (\mu\text{eV})$ &  $\Delta E_{\text{BW}} (\mu\text{eV})$	&	$\Delta E_{\text{BW}} (\mu\text{eV})$	\\
		&			& at $\hbar \omega_0 = 15$ meV	& at $\hbar \omega_0 = 10$ meV & at $\hbar \omega_0 = 5$ meV \\

\hline
$(52,0)$	&    $0.000^\circ$	& $31.100$	& $11.216$	& $1.482$	\\ 
$(53,2)$	&    $1.908^\circ$	& $31.089$	& $11.210$	& $1.466$ 	\\  
$(54,4)$	&    $3.811^\circ$	& $31.014$	& $11.193$	& $1.481$ 	\\  
$(56,9)$	&    $8.606^\circ$	& $30.732$	& $11.090$	& $1.481$	\\  
$(58,16)$	&    $15.490^\circ$	& $30.935$	& $11.162$	& $1.485$	\\  
$(59,20)$	&    $19.467^\circ$	& $31.061$	& $11.203$	& $1.461$	\\  
$(60,26)$	&    $25.598^\circ$	& $30.817$	& $11.125$	& $1.482$	\\  
$(60,28)$	&    $27.796^\circ$	& $30.962$	& $11.166$	& $1.476$	\\  
$(60,29)$	&    $28.897^\circ$	& $30.956$	& $11.176$	& $1.497$	\\  
$(60,31)$	&    $31.103^\circ$	& $30.956$	& $11.176$	& $1.498$	\\  
$(60,32)$	&    $32.204^\circ$	& $30.962$	& $11.166$	& $1.476$	\\  
$(60,34)$	&    $34.402^\circ$	& $30.817$	& $11.125$	& $1.482$	\\  
$(59,39)$	&    $40.533^\circ$	& $31.061$	& $11.203$	& $1.461$	\\  
$(58,42)$	&    $44.510^\circ$	& $30.935$	& $11.162$	& $1.484$	\\  
$(56,47)$	&    $51.394^\circ$	& $30.732$	& $11.090$	& $1.481$	\\  
$(54,50)$	&    $56.189^\circ$	& $31.014$ 	& $11.193$	& $1.481$	\\ 
$(53,51)$	&    $58.092^\circ$	& $31.089$ 	& $11.210$	& $1.465$	\\ 
$(52,52)$	&    $60.000^\circ$	& $31.100$	& $11.216$	& $1.482$	\\
\hline
\end{tabular}
%\end{ruledtabular}
\label{t: Delta E_BW example}
\end{table*}

\begin{table*}
\caption{Selected values of $\Delta E_{\text{BW}}$, in $\mu$eV, averaged over
the set of tubes of all possible chiralities $\alpha$ consistent with
the value of the radius $R\pm 0.005$ nm. The confinement energy 
$\hbar\omega_0$ is given in meV.}
%\begin{ruledtabular}
\begin{tabular}{ | c || c | c | c || c | c | c || c | c | c |}
%\hline
%\hline
%$\hbar \omega_0$ (meV) $\rightarrow$ & $15$ & $10$ & $5$ & $15$ & $10$ & $5$ & $15$ & $10$ & $5$ \\
\hline
$\epsilon$ $\rightarrow$ & $4.5$ & $4.5$ & $4.5$ & $3.5$ & $3.5$ & $3.5$ & $2.5$ & $2.5$ & $2.5$  \\
%\hline
\hline
$R$ (nm)        & $\hbar\omega_0=15$ & $\hbar\omega_0=10$  & $\hbar\omega_0=5$  & $\hbar\omega_0=15$  & $\hbar\omega_0=10$  & $\hbar\omega_0=5$  & $\hbar\omega_0=15$  & $\hbar\omega_0=10$  & $\hbar\omega_0=5$  \\
\hline
$1.018$			& $47.0$ & $17.5$ & $1.9$ &    $21.6$  &  $6.0$ & $0.4$ &      $4.5$ & $0.8$ & $< 0.1$\\
\hline
$2.036$			& $59.7$ & $23.3$ & $4.3$ &    $31.0$  & $11.2$ & $1.5$ &     $10.9$ & $3.1$ & $0.2$ \\
\hline
$2.976$			& $62.6$ & $26.0$ & $4.9$ &    $35.4$  & $13.4$ & $2.3$ &     $14.2$ & $4.7$ & $0.4$ \\
\hline
\end{tabular}
%\end{ruledtabular}
\label{t: Delta E_BW all}
\end{table*}

\section{Comparison with experiments}\label{s:exp}

So far, only one experiment \cite{Pecker13} was able to 
observe clear signatures of BW interaction in the
fine structure of the two-electron excitation spectrum. 
The evidence relied on Coulomb blockade
spectroscopy of unprecedented resolution applied to a suspended
small-gap CNT. In such device the disorder was negligible, 
as demonstrated by the substantial electron-hole symmetry of the 
measured spectrum.
It is sensible to expect further results in the near future
as a consequence of advances in device concept and 
implementation.\cite{Waissman13}

The observation of Ref.~\onlinecite{Pecker13}
builds on the comparison between the predicted energy spectrum and the
spectroscopic signal associated with the measured differential
conductance. This is a non trivial task,
as Coulomb peak positions point to the tunneling resonances
between states with one and two electrons.
In a clean sample many of these resonances turn out to be `dark',
as a consequence of the orthogonality between the states with 
one and two electrons involved in the tunneling process.
Such orthogonality is associated to either (iso)spin or orbital
degrees of freedom.\cite{Secchi12,RontaniFriedel}
For example, if the initial one-electron state has isospin $\tau=1$
and the final two-electron state has total isospin $\tau=-2$,
then the isospin blockade prevents current from flowing, 
as the isospin change in the tunneling transition $N=1\rightarrow N=2$ 
is $\Delta \tau = (-2) - (1) = -3$, which 
differs from the allowed value $\pm 1$ associated to `bright' transitions.
Another difficulty is linked to the non-equilibrium character
of the measurement, as one has to consider the metastability of
initial one-electron excited states.

With the above provisos, the following
three features of BW interaction were identified experimentally:
(i) The energy splitting $\Delta E_{\text{BW}}$ 
between the two central doublets of the S multiplet
[which is shown in column (b) of Fig.~\ref{energy levels All};
in reference \onlinecite{Pecker13} we adopted the notation
$\Delta E_{\text{VBS}} \equiv \Delta_{\text{BW}}$].
(ii) The increase of the effective spin-orbit energy splitting $\lambda$
with respect to its pristine value $\Delta E_{\text{SO}}$.
(iii) The short-range nature of BW interaction, as the states 
belonging to the AS multiplet,
which share an orbital wave function with a node, 
were unaffected by BW scattering. 

Overall, the energy structure measured in Ref.~\onlinecite{Pecker13}
was consistent with the general framework outlined in this Article and
illustrated in Fig.~\ref{energy levels All},
with the parameters 
$\hbar\omega_0\approx 8$ meV, $R\approx 3.6$ nm, and $\epsilon\approx 4.1$.
For electrons, it was found 
$\lambda = 0.40$ meV and
$\Delta E_{\text{BW}}=  -0.21 \pm 0.01$ meV, whereas 
for holes $\lambda = 0.26$ meV and
$\Delta E_{\text{BW}}=  -0.19 \pm 0.01$ meV.  
Such measured values of $\Delta E_{\text{BW}}$ are at least one
order of magnitude larger than
our predictions and have the wrong sign.
Possible drawbacks of our theory are the neglect of orbital
hybridization induced by the tube curvature
and the parametrization of Coulomb interaction through the Ohno potential. 

\section{Short-range disorder}\label{s:two_plus_disorder}

In this section we consider the effect of short-range disorder in CNTs,
as that induced by a random distribution of atomic defects. 
The scattering centers may transfer large crystal momenta to the 
conduction electrons and then mix isospins. As a consequence,
the Hamiltonian acquires a new term acting in the 
isospin space, whose effect adds to SO and BW interactions.

\subsection{Hamiltonian for short-range disorder}

We model an atomic defect at position 
${\bf{R}}^{\circledast} \equiv (R, \theta^{\circledast}, y^{\circledast})$ 
as a local single-particle scattering potential:\cite{Palyi10}
\begin{align}
\left< \vctr' \right| \hat{V}_d \left( {\bf{R}}^{\circledast} \right) \left| \vctr \right>  & = \delta\left( \vctr - \vctr'\right) V_d\left(\vctr - {\bf{R}}^{\circledast} \right) \nonumber \\
& \approx \delta\left(\vctr - \vctr'\right) V_{\delta}({\bf{R}}^{\circledast}) \delta\left( \vctr - {\bf{R}}^{\circledast}   \right) \mathcal{V}_{\text{CNT}},
\label{defect pot}
\end{align}
where $V_{\delta}$, which has the dimensions of an energy, 
is the scattering strength of the defect. This defect generates 
in the Hamiltonian the new term
\begin{align}
\hat{V}_d\left( {\bf{R}}^{\circledast} \right) = \sum_{n n'} \sum_{\sigma \sigma'} \sum_{\tau \tau'} \left< n' \sigma' \tau' \right| \hat{V}_d \left( {\bf{R}}^{\circledast}  \right) \left| n \sigma \tau \right> \hat{c}^{\dagger}_{n' \sigma' \tau'} \hat{c}_{n \sigma \tau}. 
\label{secondQ Vdef}
\end{align}
The evaluation of matrix elements 
$\left< n' \sigma' \tau' \right| \hat{V}_d \left( 
{\bf{R}}^{\circledast}  \right) \left| n \sigma \tau \right>$ 
is detailed in Appendix \ref{a:disorder}. 
The final expression of the Hamiltonian for 
a single atomic defect is
\begin{align}
&\hat{V}_d (\vctR^{\circledast}) = V_{\Delta}(\vctR^{\circledast}) \sum_{n n'} F^*_{n'}(y^{\circledast}) F_n(y^{\circledast}) \sum_{\sigma} \sum_{\tau} \nonumber \\
& \quad \quad \quad \quad \left[ \hat{c}_{n' \sigma \tau}^{\dagger} \hat{c}_{n \sigma \tau} + \frac{1}{2} \, \ee^{-\ii \tau \phi(\vctR^{\circledast})}  \hat{c}_{n' \sigma -\tau}^{\dagger} \hat{c}_{n \sigma \tau}\right] ,
\label{single imp Ham}
\end{align}
where $\phi(\vctR^{\circledast})$ is a phase that depends on the 
position $\vctR^{\circledast}$ of the atomic defect and
$V_{\Delta}(\vctR^{\circledast})=
L_y V_{\delta}({\bf{R}}^{\circledast}) / \ell_{\text{QD}}
$. 
The distribution of defects in the sample produces a sum of scattering
potentials centered at random positions  
$\vctR^{\circledast}$ 
\begin{align}
\hat{V}_d = \sum_{{\bf{R}}^{\circledast}} \hat{V}_d({\bf{R}}^{\circledast}).
\label{total imp}
\end{align}
The first-quantization analog of Eq.~\eqref{total imp} 
is expressed in terms of the axial orbital coordinate $y$ 
and isospin coordinate $t$: 
\begin{align}
\hat{V}_d(y, t) \! & = \sum_{{\bf{R}}^{\circledast}} V_{\Delta}({\bf{R}}^{\circledast})\, \delta\!\left( \frac{y - y^{\circledast}}{\ell_{\text{QD}}} \right) 
\left\{ 1 + \frac{1}{2} \left[ \ee^{\ii \phi({\bf{R}}^{\circledast})} \hat{\tau}^+(t) \right. \right. \nonumber \\
& \left. \left. + \ee^{-\ii \phi({\bf{R}}^{\circledast})} \hat{\tau}^-(t) \right]\right\}.
\label{Vdef}
\end{align}

\subsection{Short-range disorder and SO interaction 
in the one-electron system}

Similarly to the treatment of BW interaction illustrated in 
Sec.~\ref{s:two_plus_BW},
here we use first-order perturbation theory to solve the
single-particle problem in the presence of disorder. 
Therefore, assuming that the orbital excitation energies are 
larger than the splittings due to SO coupling and disorder, 
we restrict the calculation to a single orbital wave function $\psi(y)$. 

The single-particle Hamiltonian, projected on the spin-valley subspace 
of $\psi(y)$, is 
\begin{align}
\hat{H}_{\text{SP}}^{\psi} = \frac{\Delta E_{\text{SO}}}{2} \nu \hat{\sigma}  \hat{\tau} + \frac{1}{2} \left( \Delta_{d} \hat{\tau}^+ + \Delta_{d}^* \hat{\tau}^- \right),
\end{align}
with
\begin{align}
& \Delta E_{\text{SO}} \equiv 2 \Delta_{\text{SO}} \frac{\gamma}{R}, \nonumber \\
& \Delta_d \equiv \sum_{{\bf{R}}^{\circledast}} V_{\Delta}({\bf{R}}^{\circledast}) \left| \psi\left( y^{\circledast} \right) \right|^2 \ee^{\ii \phi({\bf{R}}^{\circledast})} 
\end{align}
after omitting a constant term. 
Note that the spin projection is still a good quantum number 
but the isospin is not. The two eigenvalues of the Hamiltonian 
$\hat{H}_{\text{SP}}^{\psi}$ are
\begin{align}
\pm \frac{1}{2} \sqrt{\left| \Delta_d \right|^2 + \Delta E_{\text{SO}}^2} \equiv \pm \frac{1}{2} \lambda_d, 
\end{align}
both twofold degenerate. For $-\lambda_d / 2$ the eigenstates are
\begin{align}
& \left| - \, \uparrow \right> = \chi_{+1} \otimes \frac{\Delta_d \, \varphi_{ +1} - \left( \nu \Delta E_{\text{SO}} + \lambda_d   \right) \varphi_{ -1}}{\sqrt{2 \lambda_d \left(\lambda_d + \nu \Delta E_{\text{SO}} \right)}}    , \nonumber \\        
& \left| - \, \downarrow \right> = \chi_{-1} \otimes \frac{\Delta_d^* \, \varphi_{ -1} - \left( \nu \Delta E_{\text{SO}} + \lambda_d \right) \varphi_{+1}}{\sqrt{2 \lambda_d \left(\lambda_d + \nu \Delta E_{\text{SO}} \right)}} ;
\end{align}
for $+\lambda_d / 2$ the eigenstates are
\begin{align}
& \left| + \, \uparrow \right> = \chi_{+1} \otimes \frac{\Delta_d \, \varphi_{+1} - \left( \nu \Delta E_{\text{SO}} - \lambda_d \right) \varphi_{-1}}{\sqrt{2 \lambda_d \left(\lambda_d - \nu \Delta E_{\text{SO}} \right)}} , \nonumber \\
& \left| + \, \downarrow \right> = \chi_{-1} \otimes \frac{\Delta_d^* \, \varphi_{-1} - \left( \nu \Delta E_{\text{SO}} - \lambda_d \right) \varphi_{+1}}{\sqrt{2 \lambda_d \left(\lambda_d - \nu \Delta E_{\text{SO}} \right)}} .
\end{align}
Even if each state has a non-trivial expectation value of $\hat{\tau}$ 
the sum of the expectation values for the two states of each eigenvalue 
is zero.  

\subsection{Short-range disorder, SO and BW interaction 
in the two-electron system}

Short-range disorder has important consequences for two electrons,
since it mixes states within the same orbital multiplet as well as
among multiplets of different orbital symmetries. 
In the following we consider the two limiting cases in which
the S and A multiplets are either almost degenerate or 
well separated in energy.  

The first limit occurs if the two electrons are far apart 
from each other, which can be realized either in a single quantum dot 
in the Wigner-molecule regime\cite{Secchi09,Secchi10,Secchi12,Pecker13} or 
in a double quantum dot in the $(1,1)$ charge 
configuration.\cite{Weiss10,Palyi10,vonStecher10} 
In both cases the S and A orbital multiplets are nearly degenerate and 
the orbital wave functions are well approximated by
\begin{align}
& \psi_{\text{S}}(y_1, y_2) \approx \frac{1}{\sqrt{2}} \Big[ \psi_{\text{L}}(y_1) \, \psi_{\text{R}}(y_2) + \psi_{\text{R}}(y_1) \, \psi_{\text{L}}(y_2)\Big] , \nonumber \\
& \psi_{\text{A}}(y_1, y_2) \approx \frac{1}{\sqrt{2}} \Big[ \psi_{\text{L}}(y_1) \, \psi_{\text{R}}(y_2) - \psi_{\text{R}}(y_1) \, \psi_{\text{L}}(y_2)\Big] ,
\label{WM approx}
\end{align}
where $\psi_{\text{L(R)}}(y)$ is an appropriate single-particle wave 
function centered on the left (right) classical equilibrium position.
Such position is either given by Eq.~\eqref{eq:lambda} in the WM regime or it
is the location of the QD minima in double quantum dots. 

Approximation \eqref{WM approx} holds if the overlap between $\psi_{\text{L}}$ and $\psi_{\text{R}}$ is small, $\int \psi_{\text{R}}^*(y) \psi_{\text{L}}(y) \text{d}y \approx 0$. 
In this case, each of the two electrons is sensitive only to the 
distribution of defects in the region where
its individual wave function significantly differs from zero. Therefore, it is
sufficient to solve the problem in the presence of 
defects separately for the two electrons---according to the 
procedure described in the previous subsection---and then combine 
$\psi_{\text{R}}(y)$
and $\psi_{\text{L}}(y)$ so obtained to form the two-electron eigenstates,
after Eq.~\eqref{WM approx}.

The second limit occurs if the orbital multiplets S and A are well 
separated in energy. In this case
we may apply degenerate perturbation theory separately to
the S and A multiplets, including 
disorder, SO and BW interaction, and
ignoring inter-multiplet coupling.
The matrix elements of the disorder potential 
between states with the same orbital wave function 
$\psi_k(y_1, y_2)$, with $k \in \lbrace \text{S, A} \rbrace$, are:
\begin{align}
V_d(k)_{ij} & = \delta_{i j} 2 \epsilon_d(k) + \frac{1}{2} \xi_i^{\dagger}(1, 2) \cdot \Big[ \Delta_d(k) \left( \hat{\tau}^+_1 + \hat{\tau}^+_2 \right)  \nonumber \\
& + \Delta_d^*(k) \left( \hat{\tau}^-_1 + \hat{\tau}^-_2 \right)    \Big] \cdot \xi_j(1, 2),
\label{V_d(k)_ij}
\end{align}
where $\rho_k(y^{\circledast}) \equiv \ell_{\text{QD}}^{-1} \int \left| \psi_k(y, y^{\circledast}) \right|^2 \text{d}y$, and we have defined 
\begin{align}
& \epsilon_d(k) \equiv \sum_{{\bf{R}}^{\circledast}} V_{\Delta}({\bf{R}}^{\circledast}) \rho_k(y^{\circledast}), \nonumber \\
& \Delta_d(k) \equiv \sum_{{\bf{R}}^{\circledast}} V_{\Delta}({\bf{R}}^{\circledast}) \rho_k(y^{\circledast}) \ee^{\ii \phi({\bf{R}}^{\circledast})}.
\end{align}
Below we analyze the multiplet fine structure.

\subsubsection{S multiplet}

We consider the six states of a generic S multiplet, 
written in the basis that diagonalizes $(\hat{H}_{\text{SO}} + 
\hat{V}_{\text{BW}})$. We reckon energies from $E_{\text{S}} + 
2 \epsilon_d(\text{S})$, where $E_{\text{S}}$ is the orbital energy of 
the S multiplet and $\epsilon_d(\text{S})$ is a rigid
energy shift  for all the states of the multiplet [see Eq.~\eqref{V_d(k)_ij}]. 
The disorder operator acts only on those 
states with $\sigma = 0$, that we labeled as 
$\xi^{(+)}_{0, 0}$, $\xi^{(-)}_{0, 0}$, 
$\xi^{0}_{0, +2}$, $\xi^{0}_{0, -2}$. 
On this restricted subspace the Hamiltonian, including 
disorder as well as SO and BW interaction, reads as
\begin{align}
\mathbb{H}(\text{S})^{\sigma = 0} = \left( \begin{matrix} 
\lambda                   & 0                         &  - a \Delta_d^* / 2  &  - a \Delta_d / 2   \\ 
0                         & -\lambda                  &    b \Delta_d^* / 2  &    b \Delta_d / 2   \\
- a \Delta_d / 2          & b \Delta_d / 2            &  0                   &    0                \\
- a \Delta_d^* / 2        & b \Delta_d^* / 2          &  0                   &    0           
\end{matrix}\right),
\label{H_S_0}
\end{align}
where
\begin{align}
& a \equiv \frac{\lambda - \nu \Delta E_{\text{SO}} + \Delta E_{\text{BW}}}{\sqrt{2 \lambda \left( 
\lambda - \nu \Delta E_{\text{SO}}  \right)}}, \nonumber \\
& b \equiv \frac{\lambda - \nu \Delta E_{\text{SO}} - \Delta E_{\text{BW}}}{\sqrt{2 \lambda \left( 
\lambda - \nu \Delta E_{\text{SO}} \right) }},
\end{align}
and $\lambda = \sqrt{(\Delta E_{\text{SO}})^2 + (\Delta E_{\text{BW}})^2}$, 
as in Eq.~\eqref{lambda}. All the matrix elements in \eqref{H_S_0} depend 
on the multiplet orbital wave function $\psi_{\text{S}}$. 

To proceed, we note that $a^2 + b^2 = 2$ and 
$ \lambda (b^2 - a^2) = - 2 \Delta E_{\text{BW}}$, so one eigenvalue is 
\begin{align}
E_0 = 0,
\end{align}
and the remaining three eigenvalues satisfy the following equation:
\begin{align}
E^3 - \left( \Delta E_{\text{SO}}^2 + \Delta E_{\text{BW}}^2 + \left|\Delta_d\right|^2 \right) E - \Delta E_{\text{BW}} \left|\Delta_d\right|^2 = 0.
\end{align}
This equation has one positive root, $E_{+}$, and two negative roots, 
that we call $E_<$ and $E_-$, with $E_< > E_-$. 
The complete list of the energy levels reckoned from 
$E_{\text{S}} + 2 \epsilon_d(\text{S})$, 
in decreasing order with $E_-$ being the ground state, is:
\begin{align}
& E_+ = \mathcal{L} \frac{2}{\sqrt{3}} \cos\left[ \frac{\arctan\left( \mu \right)}{3}\right],  \nonumber \\
& E_0 = 0, \nonumber \\
& E_< = - \mathcal{L} \frac{2}{\sqrt{3}} \cos\left[ \frac{\arctan\left( \mu \right) + \pi}{3}\right] ,  \nonumber \\
& E_{\text{BW}} = - \Delta E_{\text{BW}} , \nonumber \\
& E_- = - \mathcal{L} \frac{2}{\sqrt{3}} \cos\left[ \frac{\arctan\left( \mu \right) - \pi}{3}\right] ,
\label{eigenenergies}
\end{align}
with
\begin{align}
\mathcal{L} \equiv \sqrt{\Delta E^2_{\text{SO}} + \Delta E^2_{\text{BW}} + \left|\Delta_d\right|^2}
\label{Lambda}
\end{align}
and 
\begin{align}
\mu \equiv \sqrt{\frac{4 \left( \Delta E^2_{\text{SO}} + \Delta E^2_{\text{BW}} + \left|\Delta_d\right|^2\right)^3}{27 \Delta E^2_{\text{BW}} \left|\Delta_d\right|^4   } - 1} .
\label{mu}
\end{align}
The energy levels \eqref{eigenenergies} are all non degenerate but
$E_{\text{BW}}$ that is two-fold degenerate. 
In the limit of negligible disorder, 
$\left| \Delta_d \right|^2 \rightarrow 0$, one recovers the 
previous results in the presence of SO and BW interaction only, 
with $E_+ \rightarrow + \sqrt{\Delta E^2_{\text{SO}} + 
\Delta E^2_{\text{BW}}}$, $E_< \rightarrow 0$, 
$E_- \rightarrow - \sqrt{\Delta E^2_{\text{SO}} + \Delta E^2_{\text{BW}}}$.

\begin{figure}
\vspace{8mm}
\centerline{\epsfig{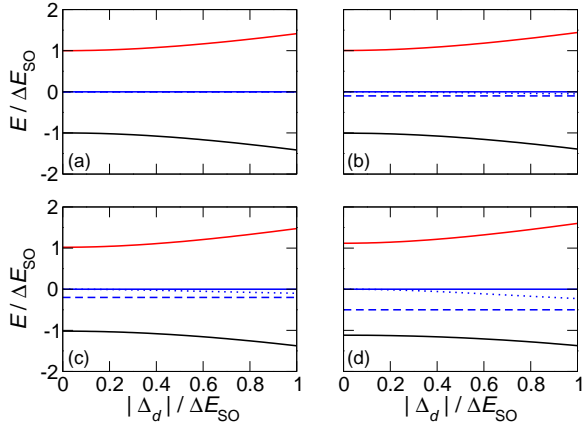}}
\caption{(Color online)  Energy fine structure of the S multiplet 
vs $\left| \Delta_d \right|$ for selected values of $\Delta E_{\text{BW}}$.
Energies are renormalized to $\Delta E_{\text{SO}}>0$.
The selected values of $\Delta E_{\text{BW}} / 
\left| \Delta E_{\text{SO}} \right|$ are: 
(a) $0$, (b) $0.1$, (c) $0.2$, (d) $0.5$. The black curve 
is $E_-$, blue dashed is $E_{\text{BW}}$, blue dotted is $E_<$, 
blue solid is $E_0 \equiv 0$, red is $E_+$. 
For $\Delta E_{\text{BW}} = 0$ [case (a)], $E_{\text{BW}} = E_< = 0 = E_0$.
\label{imp 1}}
\end{figure}

\begin{figure}
\vspace{8mm}
\centerline{\epsfig{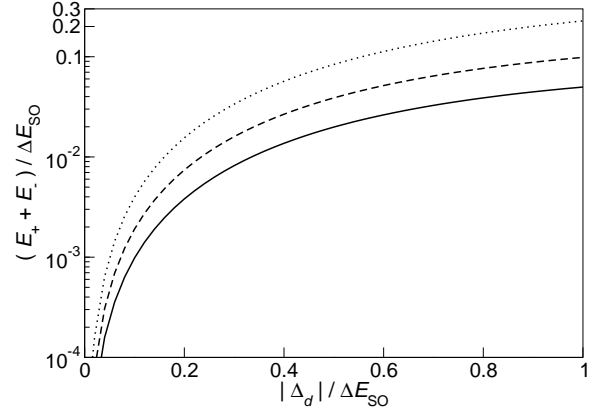}}
\caption{$(E_+ + E_-)$ vs $\left| \Delta_d \right|$ for
selected values of $\Delta E_{\text{BW}}$. 
Energies are normalized to $\Delta E_{\text{SO}} > 0$. 
The values of $\Delta E_{\text{BW}} / \left| \Delta E_{\text{SO}} \right|$ 
are: $0.1$ (solid curve), $0.2$ (dashed curve), $0.5$
(dotted curve). For either $\Delta E_{\text{BW}} = 0$ 
or $\Delta_d = 0$ one has $(E_+ + E_-) = 0$. 
\label{imp 2}}
\end{figure}

Equation \eqref{eigenenergies} shows
that the combined action of SO coupling, BW interaction, and disorder 
produces a non-trivial fine structure made
of five resolved energy levels. 
To study the dependence of the eigenvalues \eqref{eigenenergies} on 
the different contributions to isospin mixing, we consider
ultraclean devices where
disorder is a weak perturbation,\cite{Kuemmeth08} 
hence $\left| \Delta_d \right|$ is typically much smaller than 
$\left| \Delta E_{\text{SO}} \right|$.
Besides, we have 
$\left|\Delta E_{\text{BW}}\right|\ll 
\left|\Delta E_{\text{SO}}\right|$, 
hence $\left| \Delta E_{\text{SO}} \right|$ is the
dominant energy scale that we use to renormalize all energies in
Figs.~\ref{imp 1} and \ref{imp 2}.

Figures \ref{imp 1} and \ref{imp 2} show the energy levels of the S multiplet
as a function of $\left| \Delta_d \right | / 
\left| \Delta E_{\text{SO}} \right| $ for
selected values of 
$\Delta E_{\text{BW}} / \left| \Delta E_{\text{SO}} \right| $. 
Since the distribution of atomic defects is random, here we assume
the three quantities $\Delta E_{\text{SO}}$, $\Delta E_{\text{BW}}$,
and $\Delta_d$
to be uncorrelated, although they all
depend on the orbital wave function 
($\Delta_d$ through the density,
$\Delta E_{\text{BW}}$ through the pair correlation function, and 
$\Delta E_{\text{SO}}$ through the kinetic energy).

Specifically, Fig.~\ref{imp 1} focuses on the evolution of the level 
sequence vs $\Delta_d$ with the increase of BW scattering, starting from 
$\Delta E_{\text{BW}}=0$ [Fig.~\ref{imp 1}(a)].
We see that BW interaction splits the central
line at $E=0$ into two curves, whereas disorder further splits 
discernibly the zero-energy
eigenvalue when $\left| \Delta_d \right | / 
\left| \Delta E_{\text{SO}} \right|> 0.4 $ [dotted line in 
Figs.~\ref{imp 1}(c) and (d)]. 

In the absence of disorder, the energies of the two outer levels are 
symmetric with respect to $E = 0$. Disorder breaks this symmetry, 
as shown in Fig.~\ref{imp 2}. The centroid $(E_+ + E_-)$, in fact, 
deviates from zero as $\Delta_d $ increases, 
although for moderate disorder, 
say $\left| \Delta_d \right| 
\lesssim 0.2 \left| \Delta E_{\text{SO}} \right|$, 
the discrepancy is small, 
$(E_+ + E_-) \lesssim 10^{-2} \left| \Delta E_{\text{SO}} \right|$.

Putting $\Delta_d \equiv \left| \Delta_d \right| \ee^{\ii \phi}$, 
we obtain that the eigenstates associated with eigenvalues $E = 
E_{+}$, $E_{<}$, and $E_{-}$ are given by
\begin{align}
\xi_{E} & = \Bigg\{ \frac{E}{E + \Delta E_{\text{BW}}} \left[ (\lambda + E) a \, \xi_{0,0}^{(+)} + (\lambda - E) b \, \xi_{0,0}^{(-)}\right]  \nonumber \\
& \quad \quad - \left| \Delta_d\right| \left( \ee^{\ii \phi} \xi_{0, +2}^{0}  + \ee^{-\ii\phi} \xi_{0, -2}^{0}  \right) \Bigg\} \nonumber \\      
& \Bigg/ \sqrt{2 E^2 \left[ 1 + \frac{\Delta E^2_{\text{SO}}}{\left( E + \Delta E_{\text{BW}}\right)^2} \right] + 2 \left| \Delta_d\right|^2} ,
\label{eigenstates 1}
\end{align} 
the eigenstate of $E_0$ is
\begin{align}
\xi_{0} = \frac{1}{\sqrt{2}} \left( \ee^{\ii \phi} \xi_{0, +2}^0 - \ee^{-\ii \phi} \xi_{0, -2}^0 \right), 
\label{eigenstates 2}
\end{align}
and the two eigenstates corresponding to $E_{\text{BW}}$ are
\begin{align}
\xi^{\pm}_{E_{\text{BW}}} = \xi^0_{\pm 2, 0},
\label{eigenstates 3}
\end{align}
as in the absence of disorder.

\subsubsection{A multiplet}

We now consider the ten states of a generic A orbital multiplet, 
reckoning energies from $E_{\text{A}} + 2 \epsilon_d(\text{A})$. 
The disorder operator mixes states having like
spin projections $\sigma$. Since the probability for the two electrons 
to be close in space is tiny, we neglect BW scattering.\cite{mynote} 
Here we use the notation $\zeta^{\eta}_{\sigma, \tau}$ for the symmetric 
spin-valley wave functions. The sector with $\sigma = \pm 2$ consists of 
states $\zeta^{\mp 2}_{\pm 2, -2}, \zeta^{0}_{\pm 2, 0}, 
\zeta^{\pm 2}_{\pm 2, +2}$. 
In this subspace, the Hamiltonian matrix is
\begin{align}
\mathbb{H}(\text{A})^{\sigma = \pm 2} = \left( \begin{matrix} 
\mp \nu \Delta E_{\text{SO}}   &   \Delta_d^*/ \sqrt{2}   &   0                               \\ 
\Delta_d / \sqrt{2}            &   0                      &   \Delta_d^*/ \sqrt{2}            \\
0                              &   \Delta_d/ \sqrt{2}     &   \pm \nu \Delta E_{\text{SO}}    \\
\end{matrix}\right).
\end{align}
The eigenvalues are
\begin{align}
& E_+ = \sqrt{\Delta E_{\text{SO}}^2 + \left| \Delta_d \right|^2} \equiv \lambda_d, \nonumber \\
& E_0 = 0, \nonumber \\
& E_- = - \sqrt{\Delta E_{\text{SO}}^2 + \left| \Delta_d \right|^2} \equiv -\lambda_d,
\end{align}
and the corresponding eigenstates are
\begin{align}
 \zeta^{(+)}_{\sigma = \pm 2} & = \frac{1}{2 E_+ \sqrt{\gamma_{\pm}(\text{A})}} \Big[ \ee^{-\ii \phi} \gamma_{\pm}^2(\text{A}) \zeta^{\mp 2}_{\pm 2, -2}  \nonumber \\
&  + \! \sqrt{2} \left| \Delta_d \right| \left( E_+ - \nu \Delta E_{\text{SO}} \right) \zeta^{0}_{\pm 2, 0} + \left|\Delta_d\right|^2 \ee^{\ii \phi} \zeta^{\pm 2}_{\pm 2, +2} \Big] , \nonumber \\
 \zeta^{(0)}_{\sigma = \pm 2} & = \frac{1}{E_+} \Big[ \frac{1}{\sqrt{2}} \left(\Delta^*_d \zeta^{\mp 2}_{\pm 2, -2} - \Delta_d \zeta^{\pm 2}_{\pm 2, +2}\right) \nonumber \\
 & \pm \nu \Delta E_{\text{SO}} \zeta^{0}_{\pm 2, 0} \Big], \nonumber \\
 \zeta^{(-)}_{\sigma = \pm 2} & = \frac{1}{2 E_+ \sqrt{\gamma_{\mp}(\text{A})}} \Big[ \ee^{-\ii \phi} \gamma_{\mp}^2(\text{A}) \zeta^{\mp 2}_{\pm 2, -2} \nonumber \\
 & - \! \sqrt{2} \left| \Delta_d \right| \left( E_+ + \nu \Delta E_{\text{SO}} \right) \zeta^{0}_{\pm 2, 0} + \left|\Delta_d\right|^2 \ee^{\ii \phi} \zeta^{\pm 2}_{\pm 2, +2} \Big], 
\label{eigenstates 4}
\end{align}
where we have defined
\begin{align}
\gamma_{\pm}(\text{A}) \equiv \left| \Delta_d \right|^2 + 2 \Delta E_{\text{SO}} \left( \Delta E_{\text{SO}} \mp \nu E_+ \right).
\end{align}

The subspace with $\sigma = 0$ consists of 
states $\zeta^{-2}_{0, 0}, \zeta^{+2}_{0, 0}, 
\zeta^{0}_{0, -2}, \zeta^{0}_{0, +2}$. 
In this sector the Hamiltonian matrix is
\begin{align}
\mathbb{H}(\text{A})^{\sigma = 0} = \left( \begin{matrix} 
 -\nu \Delta E_{\text{SO}} &   0                         &   \Delta_d / 2    &  \Delta_d^* / 2     \\ 
 0                         &   \nu \Delta E_{\text{SO}}  &   \Delta_d / 2    &  \Delta_d^* / 2     \\ 
 \Delta_d^* / 2            &   \Delta_d^* / 2            &   0               &  0                  \\
 \Delta_d / 2              &   \Delta_d / 2              &   0               &  0                  \\
\end{matrix}\right).
\end{align}
The eigenvalues are the same as for $\sigma = \pm 2$. 
The eigenstates are:
\begin{align}
\zeta_{\sigma = 0}^{(+)} & = \frac{1}{2 \lambda_d} 
\left[ \frac{\gamma_+(\text{A}) \zeta_{0, 0}^{-2} + \left| \Delta_d \right|^2 \zeta_{0, 0}^{+2}}{\lambda_d - \nu \Delta E_{\text{SO}}} \right. \nonumber \\
& \left. + \frac{\lambda_d -\nu \Delta E_{\text{SO}}}{\sqrt{\gamma_+(\text{A})}} \left( \Delta_d^* \zeta_{0, -2}^0 + \Delta_d \zeta_{0, +2}^0 \right) \right], \nonumber \\
 \zeta_{\sigma = 0}^{(0,u)} & = \frac{1}{\sqrt{2}} \left( \ee^{-\ii \phi} \zeta_{0, -2}^0 - \ee^{\ii \phi} \zeta_{0, +2}^0 \right), \nonumber \\
 \zeta_{\sigma = 0}^{(0,g)} & = \frac{1}{\sqrt{2} \lambda_d} \left[ \left| \Delta_d \right| \left( \zeta_{0, 0}^{-2} - \zeta_{0, 0}^{+2}\right) \right. \nonumber \\
& \left. + \nu \Delta E_{\text{SO}} \left( \ee^{-\ii \phi} \zeta_{0, -2}^0 + \ee^{\ii \phi} \zeta_{0, +2}^0 \right) \right], \nonumber \\
 \zeta_{\sigma = 0}^{(-)} & = \frac{1}{2 \lambda_d} 
\left[ \frac{\gamma_-(\text{A}) \zeta_{0, 0}^{-2} + \left| \Delta_d \right|^2 \zeta_{0, 0}^{+2}}{\lambda_d + \nu \Delta E_{\text{SO}}} \right. \nonumber \\
& \left. - \frac{\lambda_d + 
\nu \Delta E_{\text{SO}}}{ \sqrt{\gamma_-(\text{A})} } 
\left( \Delta_d^* \zeta_{0, -2}^0 + \Delta_d \zeta_{0, +2}^0 \right) \right],
\label{eigenstates 5}
\end{align}
where the apex $(0,g)$ [$(0,u)$] labels the eigenspace with $E_0=0$
corresponding to the (anti)symmetric combination of states of opposite isospins
(in the absence of disorder).

\section{Conclusion}\label{s:conclusion}

In conclusion, we have provided a theory of inter-valley
scattering induced by Coulomb interaction in semiconducting 
carbon-nanotube quantum dots, which takes explicitly into account
tube chirality. Focusing on two electrons, we have 
shown that BW scattering depends on the pair correlation function 
of the interacting state. 
We have predicted previously overlooked energy splittings of the order 
of some tens of $\mu$eV in the fine structure of the lowest
symmetric orbital multiplet in typical regimes, whereas the
effect of BW interaction is negligible for antisymmetric orbital wave 
functions. 

As a by-product, we have presented analytical expressions for 
the atomic coordinates on an arbitrary CNT surface 
that depend on two indexes unrelated to the graphene geometry. 
This could be useful for evaluating microscopic 
and mechanical properties of CNTs. 

We have included in our model the effect of short-range disorder due
to a random distribution of atomic defects, being an additional
source of inter-valley scattering. 
The interplay between SO coupling, BW scattering and disorder leads to a 
rich energy spectrum for S multiplets. 
In particular, two-electron states are no more eigenstates of the 
isospin, which has implications for the studies of spin-valley blockade. 
Our findings are useful for the interpretation of Coulomb
blockade experiments in ultraclean carbon nanotubes
as well as for designing two-electron qubits. 

\section*{Acknowledgements}

This work was supported by projects CINECA-ISCRA IscrC\_TUN1DFEW, 
IscrC\_TRAP-DIP,
Fondazione Cassa di Risparmio di Modena ``COLDandFEW'',
and EU-FP7 Marie Curie Initial Training Network ``INDEX''. 
We thank Shahal Ilani, Sharon Pecker, Ferdinand Kuemmeth,
Andr\'as P\'alyi, and Guido Burkard for stimulating discussions.

\appendix
\section{Atomic coordinates}\label{a:coordinates}

In this Appendix 
we detail the derivation of Eqs.~\eqref{coordinates_nj} and
\eqref{coordinates_offsets}, starting from Eq.~\eqref{coordinates}.
Within sublattice A (B), we first determine the allowed 
axial coordinates $y_{\text{A}}$ ($y_{\text{B}}$), expressed 
as a function of the integer number $n$, and then determine the 
angular coordinate $\theta_{\text{A}}$ ($\theta_{\text{B}}$) 
of all the atoms located
on the tube circumference at $y_{\text{A}}$ ($y_{\text{B}}$).

\subsection{Axial coordinate}
The axial coordinate  $y_{\text{A}}(n_1, n_2)$
[$y_{\text{B}}(n_1, n_2)$] appearing in Eq.~\eqref{coordinates} 
may be put in one-to-one correspondence with the integer index $n$, 
varying in $(-\infty, +\infty)$ if the nanotube length is infinite. 
In order to show this we write
\begin{align}
n_a \equiv f_{ab} \nu_a, \quad n_b \equiv f_{ab} \nu_b,
\end{align}
where $f_{ab}$ is the greatest common divisor of $n_a$ and $n_b$, and the integers $\nu_a$ and $\nu_b$ are coprime.
Simplifying Eq.~\eqref{coordinates} accordingly, we obtain 
\begin{align}
& y_{\text{A}}(n_1, n_2) = y_{\text{B}}(n_1, n_2) + \Delta y_{\text{AB}}, \nonumber \\
& y_{\text{B}}(n_1, n_2) = \frac{\sqrt{3} a}{2 \sqrt{\nu^2_a + \nu^2_b - \nu_a \nu_b}} 
\Big( \nu_a n_2  - \nu_b n_1 \Big),
\label{axial coordinates}
\end{align}
where
\begin{align}
\Delta y_{\text{AB}} = \frac{\sqrt{3} a}{2 \sqrt{\nu_a^2 + \nu_b^2 - \nu_a \nu_b}} \frac{1}{3} \Big( 2 \nu_a - \nu_b\Big) 
= \frac{a}{\sqrt{3}} \cos(\alpha).
\label{offsetAB}
\end{align}
The allowed values of the axial coordinate depend on those
assumed by the integer quantity $(\nu_a n_2  - \nu_b n_1)$,
with both $n_1$ and $n_2$ belonging to 
$\mathbb{Z}$. 
To find out these values, we use
a result of discrete mathematics known as B\'ezout's lemma:\cite{Tignol01}
If $a$ and $b$ are non-zero integers 
with greatest common divisor $d$, then there exist two integers 
$n_x$ and $n_y$ such that $a n_x + b n_y = d$. 
Hence, replace $a$, $b$ and $d$,
respectively, by $\nu_a$, $\nu_b$, and $1$. 
This lemma implies that there exists a couple $(n^*_1, n^*_2)$ such that
$\nu_a n^*_2  - \nu_b n^*_1 = 1$. It follows that for any integer $n$, 
there is a couple of integers $(n_1 = n n^*_1, n_2 = n n^*_2)$ such that 
$\nu_a n_2  - \nu_b n_1 = n$. In other words, the domain for the quantity 
$(\nu_a n_2  - \nu_b n_1)$ is the whole $\mathbb{Z}$. It follows that the 
allowed axial coordinates of the carbon atoms may be written as
\begin{align}
& y_{\text{A}}(n) = y_{\text{B}}(n) + \Delta y_{\text{AB}}, \nonumber \\
& y_{\text{B}}(n) = \frac{\sqrt{3} a}{2 \sqrt{\nu^2_a + \nu^2_b - 
\nu_a \nu_b}} n,
\label{axial coordinates final}
\end{align}
for any $n \in \mathbb{Z}$.

\subsection{Angular coordinate}

We proceed to identify the atoms lying on the nanotube circumference 
at the axial coordinate $y_{\text{A}}(n)$ [$y_{\text{B}}(n)]$. 
Consider two sites of the B sublattice at
$\underline{R}_{\text{B}} \equiv \underline{R}_{\text{B}}(n_1, n_2)$ and
$\underline{R}'_{\text{B}} \equiv \underline{R}_{\text{B}}(n'_1, n'_2)$. 
From Eqs.~\eqref{axial coordinates}, 
the two sites have the same axial coordinate if
\begin{align}
y_{\text{B}} = y'_{\text{B}} \Rightarrow \nu_a (n_2 - n'_2) = 
\nu_b (n_1 - n'_1),
\label{equal_y}
\end{align}
with an analogous condition for A sublattice. 
Since $\nu_a$ and $\nu_b$ are coprime, condition \eqref{equal_y} holds
if there exists an integer $\bar{n}$ such that
\begin{align}
\left\{\begin{matrix} n_2 - n'_2 = \bar{n} \nu_b \\ n_1 - n'_1 = \bar{n} \nu_a \end{matrix} \right. . 
\label{condition n}
\end{align}
On the other hand, the difference between the two angular coordinates is:  
\begin{align}
\theta_{\text{B}} \! - \theta'_{\text{B}} & = \! \left \{ \! \pi \frac{\left( 2 \nu_a - \nu_b \right) \! \left( n_1 - n'_1 \right) + \left( 2 \nu_b - \nu_a \right) \!
\left( n_2 - n'_2 \right)}{f_{ab} \left( \nu^2_a + \nu^2_b - \nu_a \nu_b \right)} \! \right \} \nonumber \\
&\mod (2 \pi) .
\label{delta theta}
\end{align}
Substituting the condition \eqref{condition n} into \eqref{delta theta}, 
we obtain that the angular distance
between two B sites with the same axial coordinate is
\begin{align}
\Big[ \theta_{\text{B}} - \theta'_{\text{B}} \Big] \Bigg|_{y_{\text{L}} = y'_{\text{L}}} = \frac{2 \pi}{f_{ab}} \bar{n} 
\mod (2 \pi).
\end{align}
This shows that there are $f_{ab}$ distinct sublattice atoms  
on the nanotube circumference
[obtained for $\bar{n} = 0, 1, \ldots (f_{ab} - 1)$], with
the angular distance between two first neighbours
given by $2 \pi / f_{ab}$. To summarize, atoms belonging to a given sublattice
are identified by two integers: $n \in \mathbb{Z}$, 
specifying the axial coordinate $y_p(n)$ plus an angular
offset [cf.~Eq.~\eqref{coordinates_nj}], and $j \in \lbrace 0, 1, \ldots, f_{ab} - 1 \rbrace$, 
labelling the $f_{ab}$ atoms lying
on the $y = y_p(n)$ cross section.

Suppose now that the couple $(\bar{n}_1, \bar{n}_2)$ 
specifies an atom lying on the tube circumference
at $y = y_B(1)$. From Eq.~\eqref{axial coordinates}, this means that 
$\nu_a \bar{n}_2 - \nu_b \bar{n}_1 = 1$. Then, the
couple $(n_1 = n \bar{n}_1, n_2 = n \bar{n}_2)$, 
for any arbitrary $n$, specifies an atom lying
on the circumference at $y = y_B(n)$, since
$\nu_a n_2 - \nu_b n_1 = n$. Therefore, the
task of determining the allowed angular coordinates reduces to 
finding a couple
$(\bar{n}_1, \bar{n}_2)$ satisfying 
$\nu_a \bar{n}_2 - \nu_b \bar{n}_1 = 1$: once this is done, we
compute the quantity
\begin{align}
\theta_{\text{B}}(1) \equiv \pi \frac{\nu_a \left(2 \bar{n}_1 - \bar{n}_2 \right) - \nu_b \left( \bar{n}_1 - 2 \bar{n}_2\right)}{f_{ab} \left( \nu_a^2 + \nu_b^2 - \nu_a \nu_b\right)}
\label{theta_B(1)}
\end{align}
and we obtain all the angular coordinates of the atoms of the B sublattice as
\begin{align}
& \theta_{\text{B}}(\bar{n}_1, \bar{n}_2) = \theta_{\text{B}}(1) \mod (2\pi) , \nonumber \\
& \theta_{\text{B}}(n_1, n_2) = [ n \theta_{\text{B}}(1) ] \mod (2\pi),
\label{theta_B}
\end{align}
where $(n_1, n_2) = (n \bar{n}_1, n \bar{n}_2)$. Finally, the couple $(\bar{n}_1, \bar{n}_2)$ needed to evaluate $\theta_{\text{B}}(1)$ can be obtained by applying directly the extended Euclidean
algorithm. Combining Eqs.~\eqref{offsetAB}, \eqref{axial coordinates final}, \eqref{theta_B(1)} and \eqref{theta_B}, and considering the offset between A and B angular coordinates that can be evaluated directly from \eqref{coordinates}, we obtain the expressions \eqref{coordinates_nj} and \eqref{coordinates_offsets}.

\section{Single-particle states}\label{s:SP}  

In this Appendix we recall the properties of 
the eigenstates of the one-electron Hamiltonian of a quantum dot embedded
in a semiconducting carbon nanotube.

\subsection{Energy dispersion and Bloch states}

The energy dispersion $\varepsilon_{\pm}({\bf{k}})$ of graphene conduction
($+$) and valence ($-$) bands in the proximity of the non-equivalent special 
points $\bf{K}$ and $\bf{K}'$ of the first Brillouin zone
is characterized by the occurrence of Dirac cones:\cite{CastroNeto09} 
\begin{align}
\varepsilon_{\pm}({\bf{k}}) \cong \left\{ \begin{matrix} \pm \gamma \left| {\bf{k}} - {\bf{K}} \right| \quad \text{for } {\bf{k}} \approx {\bf{K}} \\ \pm \gamma \left| {\bf{k}} - {\bf{K}'} \right| \quad \text{for } {\bf{k}} \approx {\bf{K}'}\end{matrix} \right. ,
\label{Dirac cones}
\end{align}
where $\gamma \cong 533 \text{ meV}\cdot\text{nm}$ is the 
$\pi$-band parameter of graphene, 
and ${\bf{k}} = k_x \overrightarrow{\bf{x}} + k_y \overrightarrow{\bf{y}}$. 
The dispersion \eqref{Dirac cones}, due to the honeycomb 
lattice of graphene,\cite{Wallace47} may be derived most
simply by applying the tight-binding 
method, building Bloch states as superpositions 
of atomic orbitals centered on the sublattice sites.
If the orbital hybridization induced by the CNT curvature
is neglected, one may use graphene band structure to derive
the energy bands of CNTs.
This is reduced to applying a simple folding procedure
to take into account the CNT cylinder topology.\cite{DresselhausBook} 

We write the direct-space vectors lying on the CNT surface as 
${\bf{R}}_p \equiv \left( R, \underline{R}_p \right)$, 
$p \in \lbrace \text{A, B} \rbrace$, with 
$\underline{R}_p \equiv \left( \theta_p, y_p \right)$ and $R$ fixed. 
Using the azimuthal and axial coordinates, 
chiral and translation vectors are 
$\underline{L} = (2 \pi, 0)$ and $\underline{T} = (0, T)$ [note the
bar symbol labelling vectors in the $\left( \theta, y \right)$ frame]. 
We introduce generalized wave vectors of the form 
$\underline{k} \equiv \left( \kappa, k \right)$, where $\kappa$ 
is the dimensionless wave vector along the nanotube circumference 
($\kappa = k_x R$) 
and $k$ is the wave vector along the nanotube axis 
(with the dimension of the inverse of a length). 
The scalar product between $\underline{k}$ and 
generalized position vectors, of the form $\underline{r} \equiv \left( \theta, y \right)$, is defined as $\underline{k} \cdot \underline{r} = \kappa \theta + k y$. Sublattice Bloch states \cite{DresselhausBook} in CNTs are written as
\begin{align}
\psi_{p}(\underline{k}, {\bf{r}}) = \ee^{\ii \theta_p(\underline{k})} \frac{1}{\sqrt{N_c}} 
\sum_{{\bf{R}}_p} \ee^{\ii \underline{k} \cdot \underline{R}_p} 
\phi_{2 p_z}\!\!\left( {\bf{r}} - {\bf{R}}_p \right),
\end{align}
where $p \in \lbrace \text{A}, \text{B} \rbrace$, 
$\phi_{2 p_z}\!\!\left( {\bf{r}} - {\bf{R}}_p \right)$ is a single-particle 
$\pi$-band orbital centered at ${\bf{R}}_p$, 
$\theta_p(\underline{k})$ is a phase factor depending on $\underline{k}$, 
and $N_c$ is the number of sublattice sites of the CNT 
(i.e., the number of curved hexagons made of two carbon atoms each).
We assume $2 p_z$ atomic orbitals to be normalized as
\begin{align}
\int_{\text{CNT}} \left| \phi_{2 p_z}\!\!
\left( {\bf{r}} - {\bf{R}}_p \right) \right|^2 \text{d}{\bf{r}} = \mathcal{V}_{\text{CNT}},
\end{align}
where the integration is over the whole CNT and $\mathcal{V}_{\text{CNT}} = (2 \pi R) L_y L_z$, where $L_y$ is the CNT length and $L_z$ is the characteristic length associated with $2 p_z$ orbitals. \cite{Secchi10}

The proviso to include the effect of CNT curvature into the band
structure is that wave functions trasform into themselves under a 
$2 \pi$-rotation around the axis, 
$\underline{r} \rightarrow \underline{r} + \underline{L}$, 
which restricts the allowed values of the circumferential wave vector 
$\kappa$ to integer values. This condition
singles out a set of one-dimensional energy subbands, 
one subband for each value of $\kappa = n$,\cite{DresselhausBook,AndoReview} 
which are the sections of the Dirac cones at $k_x(n) = n / R$.
The intersections closest to graphene high-symmetry points 
$\bf{K}$ and $\bf{K}'$ exhibit conduction-band absolute minima 
and valence-band maxima. 
These extremal points occur at the following wave vectors:
\begin{align}
& \underline{M} = \left( \frac{n_a + n_b - \nu}{3} , \frac{n_a - n_b}{\sqrt{3} R} \right) , \nonumber \\
& \underline{M}' = \left( \frac{2 n_a - n_b + \nu}{3} , \frac{- n_b}{ \sqrt{3} R} \right) ,
\end{align}
where $\nu \in \lbrace -1, 0, +1\rbrace$ is such that $n_a + n_b = 3 n^* + \nu$, and $n^*$ is the integer closest to 
$(n_a + n_b) / 3$. The $\kappa$ components of $\underline{M}$ and $\underline{M}'$ are integer numbers (respectively, $M_{\kappa} = n^*$ and $M'_{\kappa} = n_a - n^*$), 
labelling the two one-dimensional bands in which the conduction-band minima 
lie. 

The number $\nu$ determines the electronic properties of the nanotube: 
if $\nu = 0$, the nanotube is a metal, while if $\nu = \pm 1$ the nanotube 
is a semiconductor. 
We label the two non-equivalent minima by means of the isospin index 
$\tau = +1$ $(-1)$ for point $\underline{M}$ $(\underline{M}')$. 
The Bloch states corresponding to the conduction-band minima are 
given by \cite{Secchi10}
\begin{align}
\psi_{\tau}({\bf{r}}) = \sum_{p = \text{A,B}} \! \! f^p_{\tau} \ee^{\ii \theta^p_{\tau}} \frac{1}{\sqrt{N_c}} 
\sum_{{\bf{R}}_p} \ee^{\ii \underline{M}_{\tau} \cdot \underline{R}_p} 
\phi_{2p_z} \! \left( {\bf{r}} - {\bf{R}}_p \right),
\label{SP microscopic}
\end{align}
where coefficients are given by\cite{Ando98}
\begin{align}
& f^{\text{A}}_{+1} = 1, \quad f^{\text{B}}_{+1} = \nu, \quad f^{\text{A}}_{-1} = 1, \quad f^{\text{B}}_{-1} = -\nu, 
\nonumber \\
& \theta^{\text{A}}_{+1} = 0, \quad \theta^{\text{B}}_{+1} = \alpha + 5\pi/3, 
\quad \theta^{\text{A}}_{-1} = \alpha, \quad \theta^{\text{B}}_{-1} = 0.
\end{align}
Explicitly, the CNT dispersion, close to the charge neutrality points, is:
\begin{align}
\varepsilon_{\pm}(\tau, n, k_y) = \pm \gamma \sqrt{\left[k_x(n)_{\tau}\right]^2 + k_y^2},
\label{bands nonrel}
\end{align}
where the $k_x$ and $k_y$ wave vectors are now reckoned from $\bf{K}$ and $\bf{K}'$ in valleys $\tau = +1$ and $\tau = -1$, respectively. The quantized circumferential wave vector is
\begin{align}
k_x(n)_{\tau} \equiv \frac{1}{R}\left( n - \frac{\nu \tau}{3} \right),
\end{align}
where $n \in \lbrace 0, \pm 1, \pm 2, \ldots \rbrace$.
Note the difference between the two vectors $k_x(n)_{\tau}$
and $k_x(n)=n/R$.

\subsection{Spin-orbit coupling}

Spin-orbit coupling is the first-order relativistic correction to 
the Hamiltonian and has a topological origin in CNTs due to their curvature,
as established both 
theoretically\cite{Ando00,HuertasHernando06,Zhou09,Chico09,Jeong09,Izumida09} 
and experimentally.\cite{Kuemmeth08,Churchill09,Jhang10,Jespersen11,Steele13}
Even in the absence of external magnetic
fields the combined spin and isospin fourfold degeneracy of one-electron
states is lifted, originating two Kramers doublets, each one composed of
two levels sharing the same value of the product
$\eta \equiv \sigma \tau = \pm 1$. 

Including SO, the CNT dispersion 
may be written as\cite{Ando00,HuertasHernando06,Zhou09,Chico09,Jeong09,Izumida09,Jespersen11}
\begin{align}
\varepsilon_{\pm}(\sigma, \tau, n, k_y) = \Delta_{\text{SO}}^{(0)} \cos(3 \alpha) \frac{\gamma}{R} \sigma \tau \pm \gamma \sqrt{\left[k_x(n)_{\sigma \tau}\right]^2 + k_y^2}.
\label{bands SO} 
\end{align}
Here $k_x$ depends on SO coupling,
\begin{align}
k_x(n)_{\sigma \tau} \equiv \frac{1}{R}\left( n - \frac{\nu \tau}{3} \right) - \sigma \frac{\Delta^{(1)}_{\text{SO}}}{R},
\end{align}
where $\Delta^{(0)}_{\text{SO}}$ and $\Delta^{(1)}_{\text{SO}}$ are two spin-orbit dimensionless parameters of the same order of magnitude, $\left|\Delta^{(0)}_{\text{SO}}\right| \approx \left| \Delta^{(1)}_{\text{SO}} \right| \approx 10^{-3}$. 
While the existence of the $\Delta^{(1)}_{\text{SO}}$ term was predicted 
long ago,\cite{Ando00} the occurrence of the
$\Delta_{\text{SO}}^{(0)}$ term was
proposed only recently.\cite{Zhou09,Chico09,Jeong09,Izumida09} 
The $\Delta^{(0)}_{\text{SO}}$ term has the same sign for 
electrons and holes, therefore it breaks the electron-hole symmetry, 
i.e., $\varepsilon_{-}(\sigma, \tau, n, k_y) \ne - 
\varepsilon_{+}(\sigma, \tau, n, k_y)$, consistently with 
its experimental observation.\cite{Kuemmeth08} In the following, 
we will show how the theoretical formalism that we have employed in 
our previous works\cite{Secchi09,Secchi10,Secchi12,Pecker13} 
may be extended to account 
for the new term $\Delta_{\text{SO}}^{(0)}$, as well as for the 
coupling with the axial orbital degree of freedom $y$, 
as recently observed.\cite{Jespersen11} 

\subsection{Effective mass approximation}

We now introduce a confinement potential that forms a 
1D QD, varying slowly with respect to the scale of $a$.
We focus on semiconducting CNTs, treating the 
low-energy states close to the charge neutrality point using
the envelope functions in the effective mass approximation.\cite{Luttinger55} 
This is obtained by expanding \eqref{bands SO} for $k_y^2 \ll \left[k_x(n)_{\sigma \tau}\right]^2$, retaining only terms up to order $\left[k_y /k_x(n)_{\sigma \tau}\right]^2$. Therefore,
\begin{align}
\sqrt{\left[k_x(n)_{\sigma \tau}\right]^2 + k_y^2} 
\approx 
\left| k_x(n)_{\sigma \tau} \right| + \frac{k_y^2}{2 \left| k_x(n)_{\sigma \tau} \right|}.
\end{align} 
We consider only $n = 0$, corresponding to the lowest subband 
for both electrons and holes. We have
\begin{align}
\left| k_x(0)_{\sigma \tau} \right| = \frac{1}{R} \left| \frac{\nu \tau}{3} + \Delta_{\text{SO}}^{(1)} \sigma \right| = \frac{1}{3R} + \frac{\Delta_{\text{SO}}^{(1)}}{R} \nu \sigma \tau ,
\end{align}
since $\left| \nu \tau \right| = 1$ for semiconducting CNTs ($\nu = \pm 1$), and $\left| \Delta_{\text{SO}}^{(1)} \right| \ll 1/3$. 
Therefore, the conduction and valence bands \eqref{bands SO} 
may be approximated as:
\begin{align}
\varepsilon_{\pm} \cong \pm \frac{\gamma}{3 R} + \frac{\Delta_{\text{SO} \pm} \gamma}{R} \nu \sigma \tau 
\pm \frac{1}{2} \frac{3 R \gamma}{1 + 3 \Delta_{\text{SO}}^{(1)} \nu \sigma \tau} k_y^2 ,
\label{bands SO appr}
\end{align}
where $\Delta_{\text{SO} \pm} \equiv \nu \Delta_{\text{SO}}^{(0)} \cos(3 \alpha) \pm \Delta_{\text{SO}}^{(1)}$. 
Now we make the operatorial substitution $k_y^2 \rightarrow -\partial^2 / 
\partial y^2$ and define the spin- and isospin-dependent effective mass
$m^*_{\sigma \tau}$ as 
\begin{align}
m^*_{\sigma \tau} \equiv \frac{\hbar^2}{3 R \gamma} \left( 1 + 3 \Delta_{\text{SO}}^{(1)} \nu \sigma \tau \right).
\label{effmasseta}
\end{align}
Focusing on the conduction band, 
we write the effective single-particle Hamiltonian $H_{\text{SP}}$, 
including the QD confinement potential $V_{\text{QD}}(y)$, as 
\begin{align}
H_{\text{SP}} = \frac{\gamma}{3 R} + \frac{\Delta_{\text{SO} +} \gamma}{R} \nu \sigma \tau - \frac{\hbar^2}{2 m^*_{\sigma \tau}} \frac{\partial^2}{\partial y^2} + V_{\text{QD}}(y).
\label{H_SP nonappr}
\end{align}
We note from Eq.~\eqref{effmasseta} that the spin- and isospin- dependent part of $m^*_{\sigma \tau}$ is of the order of $\approx 10^{-3} m^*$, where $m^*$ is the orbital effective mass,
\begin{align}
m^* \equiv \frac{\hbar^2}{3 R \gamma}.
\label{effmass}
\end{align}
Therefore, we can expand $1 / m^*_{\sigma \tau}$ in Eq.~\eqref{H_SP nonappr} to the first order in $\Delta_{\text{SO}}^{(1)}$, obtaining
\begin{align}
H_{\text{SP}} \cong & \frac{\gamma}{3 R} + \nu \sigma \tau 
\left( \frac{\Delta_{\text{SO} + } \gamma}{R} + 3 \Delta_{\text{SO}}^{(1)} \frac{\hbar^2}{2 m^*} \frac{\partial^2}{\partial y^2}\right) \nonumber \\
& - \frac{\hbar^2}{2 m^*} \frac{\partial^2}{\partial y^2} + V_{\text{QD}}(y).
\label{H_SP}
\end{align}
In the case of a gate-defined QD embedded in a CNT the 
confinement potential is parabolic:
\begin{align}
V_{\text{QD}}(y) = \frac{1}{2} m^* \omega_0^2 y^2,
\end{align}
with $\omega_0$ being the characteristic harmonic oscillator frequency.
The QD size in real space is given by the characteristic 
length $\ell_{\text{QD}} = \sqrt{\hbar / (m^* \omega_0)}$.

A QD single-particle state is the product of 
the microscopic Bloch function \eqref{SP microscopic} 
times the slowly-varying envelope function $F_{n \tau \sigma}(y)$ 
times the spinor $\chi_{\sigma}$,
\begin{align}
\psi_{n \tau \sigma}({\bf{r}}, s) = \mathcal{N} F_{n \sigma \tau}(y) \psi_{\tau}({\bf{r}}) \chi_{\sigma}(s),
\label{SP functions}
\end{align}
where $\mathcal{N}$ is a normalization constant that will be specified later. 
The envelope function $F_{n \sigma \tau}(y)$ satisfies 
the eigenvalue equation
\begin{align}
H_{\text{SP}} F_{n \tau \sigma} = \epsilon_{n \tau \sigma} F_{n \tau \sigma},
\end{align}
with $n$ labelling the orbital states. 
For simplicity, we use again first-order perturbation theory, 
assuming that the orbital functions do not depend on $\sigma$ and $\tau$. 
Therefore, we assume that $F_n$ solves the eigenvalue problem 
\begin{align}
\left[ - \frac{\hbar^2}{2 m^*} \frac{\partial^2}{\partial y^2} + V_{\text{QD}}(y) \right] F_n(y) = \epsilon_n F_n(y),
\end{align}
where $\epsilon_n$ is a QD discrete level 
and the envelope function normalization is
\begin{align}
\int_{-\infty}^{+ \infty} F^*_n(y) F_{n'}(y) \text{d}y = \ell_{\text{QD}} \delta_{n, n'}.
\end{align}

The orbital envelope function can be multiplied by the four 
microscopic states $\psi_{\tau}({\bf{r}}) \chi_{\sigma}(s)$, with
$\tau \in \lbrace +1, -1 \rbrace$ and $\sigma \in \lbrace +1, -1 \rbrace$. 
Each one of these states has energy $\varepsilon_{n \tau \sigma}$, with
\begin{align}
\varepsilon_{n \tau \sigma} & = \frac{\gamma}{3 R} + \varepsilon_n + \left( \frac{\Delta_{\text{SO} + } \gamma}{R} - 3 \Delta_{\text{SO}}^{(1)} \left < E_{n}^{\text{kin}} \right > \right) \nu \sigma \tau \nonumber \\
& \equiv \frac{\gamma}{3 R} + \varepsilon_n + \frac{\Delta_{\text{SO}} \gamma}{R} \nu \sigma \tau, 
\label{SO_disp}
\end{align}
and
\begin{align}
\left < E_{n}^{\text{kin}} \right > = - \frac{\hbar^2}{2 m^*} \ell_{\text{QD}}^{-1} \int_{- \infty}^{+ \infty} F^*_n(y) \frac{\partial^2}{\partial y^2} F_n(y) \text{d}y
\label{kin correction}
\end{align}
is a kinetic-energy correction to the effective spin-orbit parameter,
\begin{align}
\Delta_{\text{SO}} \equiv \Delta_{\text{SO} + } - 3 \Delta_{\text{SO}}^{(1)} \frac{ R \left < E_{n}^{\text{kin}} \right >}{\gamma}.
\label{eff SO}
\end{align}
According to Eq.~\eqref{SO_disp}, the single-particle orbital 
energy $\varepsilon_{n \tau \sigma}$ is split into two levels, 
corresponding respectively to $\sigma \tau = + 1$ and $\sigma \tau = -1$. 
The gap between such states is $2 \Delta_{\text{SO}} \gamma / R$, where, as shown by Eq.~\eqref{eff SO}, the SO parameter $\Delta_{\text{SO}}$ varies with the orbital multiplet under consideration through the kinetic term \eqref{kin correction}. This effect has also been observed experimentally: \cite{Jespersen11} in particular, it has been shown that the SO gap changes as the confinement potential is modified by an electrostatic gate. In the following we will just take $\Delta_{\text{SO}}$ as a parameter, keeping in mind that it can change in magnitude and sign with the orbital multiplet under consideration.

Finally, the normalization constant $\mathcal{N}$ in Eq.~\eqref{SP functions} is evaluated with a procedure \cite{Secchi10} that exploits the localization of the atomic orbitals $\phi_{2 p_z}$ appearing in the Bloch states \eqref{SP microscopic} around the positions of the respective carbon nuclei; the result is 
\begin{align}
\mathcal{N} = \frac{1}{2 \sqrt{\pi R L_z \ell_{\text{QD}}}},
\end{align}
providing the following normalization of the orbital wave functions:
\begin{align}
\mathcal{N}^2 \int_{\text{CNT}} \left[ F^*_n(y) \psi^*_{\tau}({\bf{r}}) \right] \left[ F_{n'}(y) \psi_{\tau'}({\bf{r}}) \right] \text{d}{\bf{r}} = \delta_{n, n'} \delta_{\tau, \tau'}.
\end{align}

\section{The BW scattering term of the Hamiltonian}\label{a:BW}

In this Appendix we detail the passages that lead from the 
expression \eqref{BW_element} to the form \eqref{BW matrix element} 
of BW scattering matrix elements. 

Using Eqs.~\eqref{U dependence} and \eqref{Delta M}, 
we write \eqref{BW_element} as
\begin{align}
& V^{(\text{BW})}_{a, b; c, d}(\tau)  = \frac{L_y^2}{4 N_c^2} \ell_{\text{QD}}^{-2} \sum_{p,p'} 
\ee^{\ii \tau \phi_{p p'}} \sum_{n \in \mathbb{Z}} \sum_{j = 0}^{f_{ab} - 1} 
\sum_{n' \in \mathbb{Z}} \sum_{j' = 0}^{f_{ab} - 1} \nonumber \\
& \ee^{\ii \tau \Delta M_{\kappa} \cdot \left[ \theta_p(n,j) - \theta_{p'}(n',j')\right]} 
\ee^{\ii \tau \Delta M_k \cdot \left[ y_p(n) - y_{p'}(n')\right]} \nonumber \\
&  \times U\! \left\{ \left[ y_p(n) - y_{p'}(n')\right]^2, \sin^2 \! \left[ \frac{\theta_p(n,j) - \theta_{p'}(n',j')}{2}\right] 
\right\} \nonumber \\
&  \times F_{a}^*[y_p(n)] F_{b}^*[y_{p'}(n')] F_{c}[y_{p'}(n')] F_{d}[y_{p}(n)].
\label{BW_element_bis}
\end{align}
To simplify this expression, we note that it contains both quantities 
that vary slowly (the envelope functions) and quantities that vary rapidly 
(the exponentials and the interaction potential) with respect to the 
indexes $n$ and $n'$ labelling the axial coordinates. Nevertheless, 
the dependence of the rapidly-varying quantities on the axial coordinates 
occurs through the difference
\begin{align}
y_p(n) - y_{p'}(n') = \Delta y_{p p'} + y_{\text{B}}(n-n'),
\label{diff y}
\end{align}
with $\Delta y_{\text{BA}} = - \Delta y_{\text{AB}}$, and $\Delta y_{\text{AA}} = \Delta y_{\text{BB}} = 0$, hence the expression on the
left hand side of \eqref{diff y} depends only on $(n - n')$. 

Now consider the dependence of the expression \eqref{BW_element_bis}
on the angular coordinates, namely on the quantity 
$\theta_p(n,j) - \theta_{p'}(n',j')$. This must be investigated 
with some care. Below we show that, if $p = p'$, the angular
difference is an angle pointing to the B sublattice, 
otherwise it points to the A sublattice. 
This is seen most easily by considering the representation of 
angles in terms of indexes $(n_1, n_2)$, 
Eq.~\eqref{coordinates}. 
Specifically, given $(n,j)$ and $(n',j')$, 
consider two couples $(\bar{n}_1, \bar{n}_2)$ 
and $(\bar{n}'_1, \bar{n}'_2)$ such that
\begin{align}
&y_p(\bar{n}_1, \bar{n}_2) = y_p(n), \quad \theta_p(\bar{n}_1, \bar{n}_2) = \theta_p(n,j), \nonumber \\
&y_{p'}(\bar{n}'_1, \bar{n}'_2) = y_{p'}(n'), \quad \theta_{p'}(\bar{n}'_1, \bar{n}'_2) = \theta_{p'}(n',j').
\end{align}
We need to evaluate
\begin{align}
\theta_p(n,j) - \theta_{p'}(n',j') = \theta_p(\bar{n}_1, \bar{n}_2) - \theta_{p'}(\bar{n}'_1, \bar{n}'_2).
\label{general theta}
\end{align} 
We distinguish two possibilities: $p = p'$ and $p \ne p'$. 

We first show that, if $p = p'$, \eqref{general theta} is equal to 
\begin{align}
\theta_p(\bar{n}_1, \bar{n}_2) - \theta_p(\bar{n}'_1, \bar{n}'_2) = \theta_{\text{B}}(\bar{n}_1 - \bar{n}'_1, \bar{n}_2 - \bar{n}'_2) + 2 \pi m
\end{align}
for some integer $m$. 
Certainly, one of the axial coordinates consistent with the angle 
$\theta_{\text{B}}(\bar{n}_1 - \bar{n}'_1, \bar{n}_2 - \bar{n}'_2)$ is
\begin{align}
y_{\text{B}}(\bar{n}_1 - \bar{n}'_1, \bar{n}_2 - \bar{n}'_2) & = y_{p}(\bar{n}_1, \bar{n}_2) - y_{p}(\bar{n}'_1, \bar{n}'_2) \nonumber \\
& = y_{p}(n) - y_{p}(n') = y_{\text{B}}(n-n'). 
\end{align}
This means that there exist integers $j^*$ and $q$, depending on $j$ and $j'$, such that
\begin{align}
\theta_p(n,j) - \theta_p(n',j') = \theta_{\text{B}}[n-n', j^*(j,j')] + 2 \pi q(j,j'),
\label{j^*}
\end{align}
where $j^*$ is not necessarily equal to $j - j'$, but it can be 
constrained to lie in the interval $\lbrace 0, \ldots, f_{ab}-1 \rbrace$ 
by means of an appropriate choice of $q$. 
Since we must evaluate the following double sum in \eqref{BW_element_bis},
\begin{align}
v_{pp} \equiv & \sum_{j = 0}^{f_{ab} - 1} \sum_{j' = 0}^{f_{ab} - 1} \ee^{\ii \tau \Delta M_{\kappa} \cdot \left[ \theta_p(n,j) - \theta_{p}(n',j')\right]} \nonumber \\ 
& \times U\left\{ \left[ y_{\text{B}}(n-n') \right]^2, \sin^2\left[ \frac{\theta_p(n,j) - \theta_p(n',j')}{2}\right] \right\} ,
\label{quantity jj'}
\end{align}
it is easy to see that, for every fixed $j$, 
as $j'$ varies in $\lbrace 0, \ldots, f_{ab}-1 \rbrace$, 
the quantity $j^*(j,j')$ in eq.~\eqref{j^*} may be let to vary in 
the same interval $\lbrace 0, \ldots, f_{ab}-1 \rbrace$ by 
appropriately choosing the integers $q(j,j')$. 
The key observation is that the actual values of $q(j,j')$ do 
not matter in the evaluation of eq.~\eqref{quantity jj'}, 
since $\Delta M_{\kappa}$ is an integer number and the interaction potential 
is periodic in the angular coordinates. So, the 
quantity \eqref{quantity jj'} is equal to:
\begin{align}
v_{pp} = & f_{ab} \sum_{j = 0}^{f_{ab} - 1} \ee^{\ii \tau \Delta M_{\kappa} \cdot \theta_{\text{B}}(n-n',j)} \nonumber \\
& \times U\left\{ \left[ y_{\text{B}}(n-n') \right]^2, \sin^2\left[ \frac{\theta_{\text{B}}(n-n',j)}{2}\right] \right\}.
\end{align}

We now consider Eq.~\eqref{general theta} for $p \ne p'$. 
If $p = \text{A}$ and $p' = \text{B}$, we obtain 
\begin{align}
\theta_{\text{A}}(\bar{n}_1, \bar{n}_2) - \theta_{\text{B}}(\bar{n}'_1, \bar{n}'_2) = \theta_{\text{A}}(\bar{n}_1 - \bar{n}'_1, \bar{n}_2 - \bar{n}'_2) + 2 \pi m
\end{align}
for some integer $m$. Similarly to the case $p = p'$ 
it can be shown that this angle is consistent with 
$y_{\text{A}}(n-n')$, and we obtain
\begin{align}
v_{AB} & \equiv  \sum_{j = 0}^{f_{ab} - 1} \sum_{j' = 0}^{f_{ab} - 1} \ee^{\ii \tau \Delta M_{\kappa} \cdot \left[ \theta_{\text{A}}(n,j) - \theta_{\text{B}}(n',j')\right]} \nonumber \\
& \times U\left\{ \left[ y_{\text{A}}(n-n') \right]^2, \sin^2\left[ \frac{\theta_{\text{A}}(n,j) - \theta_{\text{B}}(n',j')}{2}\right] \right\} \nonumber \\
&= f_{ab} \sum_{j = 0}^{f_{ab} - 1} \ee^{\ii \tau \Delta M_{\kappa} \cdot \theta_{\text{A}}(n-n',j)} \nonumber \\
& \times U\left\{ \left[ y_{\text{A}}(n-n') \right]^2, \sin^2\left[ \frac{\theta_{\text{A}}(n-n',j)}{2}\right] \right\}.
\end{align} 
Analogously, if $p = \text{B}$ and $p' = \text{A}$, we can write
\begin{align}
\theta_{\text{B}}(\bar{n}_1, \bar{n}_2) - \theta_{\text{A}}(\bar{n}'_1, \bar{n}'_2) = -\theta_{\text{A}}(\bar{n}'_1 - \bar{n}_1, \bar{n}'_2 - \bar{n}_2) + 2 \pi m.
\end{align}
This is similar to the case $(p,p') = (\text{A} ,\text{B})$ 
by exchanging $n$ with $n'$ and $j$ with $j'$ in the corresponding term 
of Eq.~\eqref{BW_element_bis}. 

We next combine the above results for $v_{pp'}$ 
with the definitions of $U_p(n,j)$, 
$f_{\tau}(n)$, and $g_{\tau}(n)$, given respectively in Eqs.~\eqref{U_p} and 
\eqref{f and g},
to rewrite the BW scattering matrix elements \eqref{BW_element_bis} as
\begin{align}
& V^{(\text{BW})}_{a, b; c, d}(\tau) = \frac{L_y^2}{4 N_c^2} \ell_{\text{QD}}^{-2}  
f_{ab} \sum_{n \in \mathbb{Z}} \sum_{n' \in \mathbb{Z}}  \Bigg\{ \nonumber \\
& f_{\tau}(n) \left\{ F_{a}^*[y_{\text{A}}(n + n')] F_{b}^*[y_{\text{A}}(n')] F_{c}[y_{\text{A}}(n')] \right. \nonumber \\ 
& \times F_{d}[y_{\text{A}}(n + n')] + F_{a}^*[y_{\text{B}}(n + n')] F_{b}^*[y_{\text{B}}(n')] \nonumber \\ 
& \left. \times F_{c}[y_{\text{B}}(n')] F_{d}[y_{\text{B}}(n + n')] \right\} \nonumber \\
& + g_{\tau}(n) F_{a}^*[y_{\text{A}}(n + n')] F_{b}^*[y_{\text{B}}(n')] F_{c}[y_{\text{B}}(n')] \nonumber \\ 
& \times F_{d}[y_{\text{A}}(n + n')] + g_{-\tau}(n) F_{a}^*[y_{\text{B}}(n')] F_{b}^*[y_{\text{A}}(n + n')] \nonumber \\
& \times F_{c}[y_{\text{A}}(n + n')] F_{d}[y_{\text{B}}(n')] \Bigg\}.
\label{BW_element simplified again}
\end{align}
We put $y_{\text{A}}(n + n') = y_{\text{A}}(n') + y_{\text{B}}(n)$ 
in the first addendum, 
$y_{\text{B}}(n + n') = y_{\text{B}}(n') + y_{\text{B}}(n)$ 
in the second addendum, 
$y_{\text{A}}(n + n') = y_{\text{B}}(n') + y_{\text{A}}(n)$ 
in the third and fourth addenda, and observe that the quantities 
varying rapidly along the axis depend only on $n$, 
whereas the coordinates depending on $n'$ appear only as arguments 
of the slowly varying envelope functions. 
Therefore, we evaluate the sum over $n'$ in the continuum limit as an 
integral,
\begin{align}
f_{ab} \sum_{n' \in \mathbb{Z}} w[y_p(n')] \approx \frac{N_c}{L_y} \int^{+\infty}_{-\infty} w(y') \text{d} y',
\end{align}
where $N_c$ is the number of lattice sites of the CNT, i.e., 
the number of atoms of each sublattice $p =$ A or B. 
Using the identity $L_y / N_c = (\sqrt{3}a^2/2)  / (2 \pi R)$, where 
$\sqrt{3}a^2/2$ is the area of the graphene unit cell, 
we eventually obtain Eq.~\eqref{BW matrix element}.

\section{Matrix elements for short-range disorder}\label{a:disorder}

In this Appendix we derive the matrix elements of the 
single-particle disorder Hamiltonian \eqref{secondQ Vdef}. 
Considering both the envelope functions and Bloch states, 
the matrix element between single-particle states is 
\begin{align}
 \left< n' \sigma' \tau' \right| & \hat{V}_d\left( {\bf{R}}^{\circledast}   \right) \left| n \sigma \tau \right>  = \delta_{\sigma \sigma'} V_{\delta}({\bf{R}}^{\circledast}) \nonumber \\
& \times \! \left[ \ell_{\text{QD}}^{-1} F^*_{n'}(y^{\circledast}) F_n(y^{\circledast}) \right] \left[ \mathcal{V}_{\text{CNT}} \psi^*_{\tau'}({\bf{R}}^{\circledast}) \psi_{\tau}({\bf{R}}^{\circledast})    \right] . 
\end{align}
Expanding the Bloch states $\psi_{\tau}({\bf{R}}^{\circledast})$ 
over the localized $2p_z$ orbitals and neglecting off-diagonal
contributions, the above expression is turned into
\begin{align}
& \left< n' \sigma' \tau' \right| \hat{V}_d ({\bf{R}}^{\circledast}) \left| n \sigma \tau \right> \approx \delta_{\sigma \sigma'} \frac{V_{\Delta}({\bf{R}}^{\circledast})}{2 N_c }  F_{n'}^*(y^{\circledast}) F_n(y^{\circledast}) \nonumber \\
& \times \! \sum_{p} \! f^p_{\tau'} f^p_{\tau} \ee^{\ii (\theta^p_{\tau} - \theta^p_{\tau'})} \! \sum_{\lbrace \vctR_p\rbrace} \! \ee^{\ii \left( \underline{M}_{\tau} - \underline{M}_{\tau'} \right)\cdot \underline{R}_p}
  \! \left| \phi_{2p_z} \! \left( {\bf{R}}^{\circledast} - \vctR_p \right) \right|^2,
\label{GeneralMtr}
\end{align}
with $V_{\Delta}({\bf{R}}^{\circledast}) \equiv L_y V_{\delta}({\bf{R}}^{\circledast}) / \ell_{\text{QD}}$.

We now distinguish two cases: $\tau' = \tau$ (isospin conserved) and 
$\tau' = - \tau$ (isospin flipped). 
For $\tau' = \tau$, Eq.~\eqref{GeneralMtr} becomes 
\begin{align}
\left< n' \sigma' \tau \right| \hat{V}_d\left( {\bf{R}}^{\circledast}  \right) \left| n \sigma \tau \right> & = \delta_{\sigma \sigma'}  \frac{V_{\Delta}({\bf{R}}^{\circledast})}{2 N_c}  F_{n'}^*(y^{\circledast}) F_n(y^{\circledast}) \nonumber \\
& \times \sum_{\lbrace \vctR \rbrace} \left| \phi_{2p_z} \left( {\bf{R}}^{\circledast} - \vctR \right) \right|^2 ,
\end{align} 
where $\lbrace \vctR \rbrace = \lbrace \vctR_A \rbrace \bigcup \lbrace \vctR_B \rbrace$ is the set of all atomic positions. We evaluate the summation over $\lbrace \vctR \rbrace$ in the continuum limit,
\begin{align}
\sum_{\lbrace \vctR \rbrace} w\left(\vctr - \vctR \right) \approx 
\left(\frac{\Delta N}{\Delta \mathcal{V}}\right) \int_{\NT} w\left(\vctr - \vctR \right) {\text{d}} R^{(3)}, 
\end{align}
where 
\begin{equation*}
\left(\frac{\Delta N}{\Delta \mathcal{V}}\right) = \frac{2 N_c}{\mathcal{V}_{\text{CNT}}}
\end{equation*}
is the density of atomic sites. The localization of $2p_z$-orbitals is 
exploited by adopting the usual approximation:
\begin{align}
\left| \phi_{2p_z} \left( \vctR^{\circledast} - \vctR \right) \right|^2 \approx \delta\left( \vctR^{\circledast} - \vctR \right) 
\mathcal{V}_{\text{CNT}},
\end{align}
which is consistent with the chosen normalization of the atomic 
orbitals.\cite{Secchi10} The result is:
\begin{align}
\left< n' \sigma' \tau \right| \hat{V}_d (\vctR^{\circledast}) \left| n \sigma \tau \right> 
= \delta_{\sigma \sigma'}  V_{\Delta}({\bf{R}}^{\circledast}) F_{n'}^*(y^{\circledast}) F_n(y^{\circledast}) .
\label{isospin cons}
\end{align}

In the case $\tau' = -\tau$, the matrix element \eqref{GeneralMtr} 
evaluated for $\tau = -1$ is equal to the complex conjugate of the matrix element evaluated for $\tau = +1$. The latter is given by 
\begin{align}
& \left< n', \sigma', -1 \right| \hat{V}_d (\vctR^{\circledast}) \! \left| n, \sigma, +1 \right> \nonumber \\
& = \delta_{\sigma, \sigma'}  \frac{V_{\Delta}({\bf{R}}^{\circledast})}{2 N_c}  F_{n'}^*(y^{\circledast}) F_n(y^{\circledast}) \sum_{p} f^p_{-1} f^p_{+1} \ee^{\ii (\theta^p_{+1} - \theta^p_{-1})} \nonumber \\ 
 & \times \sum_{\lbrace \vctR_p\rbrace} \! \ee^{\ii \left( \underline{M} - \underline{M}' \right)\cdot \underline{R}_p} \! \left| \phi_{2p_z} \left( \vctR^{\circledast} - 
\vctR_p \right) \right|^2
\end{align}
We evaluate the lattice summation in the continuum limit (the density of sublattice atoms is $N_c / \mathcal{V}_{\NT}$), obtaining
\begin{align}
& \left< n', \sigma', -1 \right| \hat{V}_d (\vctR^{\circledast}) \left| n, \sigma, +1 \right> \nonumber \\
&  = \delta_{\sigma, \sigma'}  \frac{V_{\Delta}({\bf{R}}^{\circledast})}{2} F_{n'}^*(y^{\circledast}) F_n(y^{\circledast})  \ee^{\ii \left(\underline{M} - \underline{M}'\right) \cdot \vctR^{\circledast} } \nonumber \\
& \quad  \times \left\{ \ee^{- \ii \alpha} \Theta\left(\vctR^* \in \lbrace \vctR_A \rbrace\right)
+ \ee^{\ii \alpha} \ee^{\ii \frac{2 \pi}{3}} \Theta\left(\vctR^* \in \lbrace \vctR_B \rbrace \right) \right\}
\nonumber \\
& \equiv \delta_{\sigma, \sigma'} \frac{V_{\Delta}({\bf{R}}^{\circledast})}{2}  F_{n'}^*(y^{\circledast}) F_n(y^{\circledast}) \ee^{-\ii \phi\left(\vctR^{\circledast}\right)},
\label{isospin scatt}
\end{align}
where $\Theta(x)=1$ if the argument $x$ is true otherwise $\Theta(x)=0$
and $\phi\left(\vctR^{\circledast}\right)$ is a phase factor dependent on 
the specific position of the atomic defect. Combining 
Eqs.~\eqref{isospin cons} and \eqref{isospin scatt}, one obtains the 
total Hamiltonian for an atomic defect at position 
${\bf{R}}^{\circledast}$, Eq.~\eqref{single imp Ham}.


\begin{thebibliography}{99}
\bibitem{mynote} We verified that for A multiplets the splitting $\Delta E_{\text{BW}}$ between states with $\eta = 0$ is of the order of $\approx 10^{-4} \,\mu$eV, therefore completely negligible.
\end{thebibliography}
\end{document}